\documentclass[prl,superscriptaddress,notitlepage,twocolumn,aps,longbibliography]{revtex4-1}
\usepackage{color}
\usepackage{graphicx}
\usepackage{bm}
\usepackage{amsmath}
\usepackage{natbib}
\usepackage{hyperref}

\newcommand{\sgn}{\mbox{\sf sgn}}
\newcommand{\Hcii}{H_{\text{c2}}}
\newcommand{\Hm}{H_{\text m}}

\newcommand{\UPt}{UPt$_3$}

\newcommand{\hex}{D$_{6h}$}

\begin{document}
\title{Vortex lattices and broken time reversal symmetry in the topological superconductor {\UPt}}

\author{K.~E.~Avers}
\affiliation{Department of Physics and Astronomy, Northwestern University, Evanston, IL 60208, USA}

\author{W.~J.~Gannon}
\altaffiliation{Current address: Stewart Blusson Quantum Matter Institute, University of British Columbia, Vancouver, BC V6T 1Z4, Canada}
\affiliation{Department of Physics and Astronomy, Northwestern University, Evanston, IL 60208, USA}

\author{S.~J.~Kuhn}
\altaffiliation{Current address: Center for Exploration of Energy \& Matter, Indiana University, Bloomington, IN 47408, USA}
\affiliation{Department of Physics, University of Notre Dame, Notre Dame, IN 46556, USA}

\author{W.~P.~Halperin}
\affiliation{Department of Physics and Astronomy, Northwestern University, Evanston, IL 60208, USA}

\author{J.~A.~Sauls}
\altaffiliation{Address: Department of Physics and Astronomy, Northwestern University, Evanston, IL 60208, USA}
\affiliation{Center for Applied Physics \& Superconducting Technologies, Northwestern University, Evanston, IL 60208, USA}

\author{L.~DeBeer-Schmitt}
\affiliation{Oak Ridge National Laboratory, Oak Ridge, TN 37831, USA}

\author{C.~D.~Dewhurst}
\affiliation{Institut Laue-Langevin, 71 avenue des Martyrs, CS 20156, F-38042 Grenoble cedex 9, France}

\author{J.~Gavilano}
\affiliation{Laboratory for Neutron Scattering, Paul Scherrer Institute, CH-5232 Villigen, Switzerland}

\author{G.~Nagy}
\altaffiliation{Current address: MTA Wigner Research Centre for Physics, H-1121 Budapest, Hungary}
\affiliation{Laboratory for Neutron Scattering, Paul Scherrer Institute, CH-5232 Villigen, Switzerland}

\author{U.~Gasser}
\affiliation{Laboratory for Neutron Scattering, Paul Scherrer Institute, CH-5232 Villigen, Switzerland}

\author{M.~R.~Eskildsen}
\email{Corresponding author: eskildsen@nd.edu}
\affiliation{Department of Physics, University of Notre Dame, Notre Dame, IN 46556, USA}

\begin{abstract}
The topological superconductor {\UPt}, has three distinct vortex phases, a strong indication of its unconventional character.
Using small-angle neutron scattering we have probed the vortex lattice in the {\UPt} B phase with the magnetic field along the crystal $c$-axis.
We find a difference in the vortex lattice configuration depending on the sign of the magnetic field relative to the field direction established upon entering the B phase at low temperature in a field sweep, showing that the vortices in this material posses an internal degree of freedom.
This observation is facilitated by the discovery of a field driven non-monotonic vortex lattice rotation, driven by competing effects of the superconducting gap distortion and the vortex-core structure.
From our bulk measurements we infer that the superconducting order parameter in the {\UPt} B phase breaks time reversal symmetry and exhibits chiral symmetry with respect to the $c$-axis.
\end{abstract}

\date{\today}

\maketitle

\section{I. Introduction}
Topological properties of materials are of fundamental as well as practical importance, and are currently the focus of intense research~\cite{Qi:2011hba}. Recently, topological superconductors have received increased attention, as it was realized that vortices in them can host Majorana fermions with potential use in quantum computing~\cite{Schnyder:2015kt,Beenakker:2016ds,Kozii:2016go}. As a result, some well known unconventional superconductors have received renewed attention~\cite{Mizushima:2016hk,Sato:2017th}, particularly {\UPt}~\cite{Strand:2009eq,Schemm:2014fv}, Sr$_2$RuO$_4$~\cite{Maeno:2012ew,Steppke:2017fb} and superfluid $^3$He~\cite{Ikegami:2013jx,Autti:2016bu}. These materials are believed to posses chiral ground states with broken mirror and time-reversal symmetries, as well as chiral fermions associated with the topology of the chiral ground state.

\begin{figure}
\includegraphics{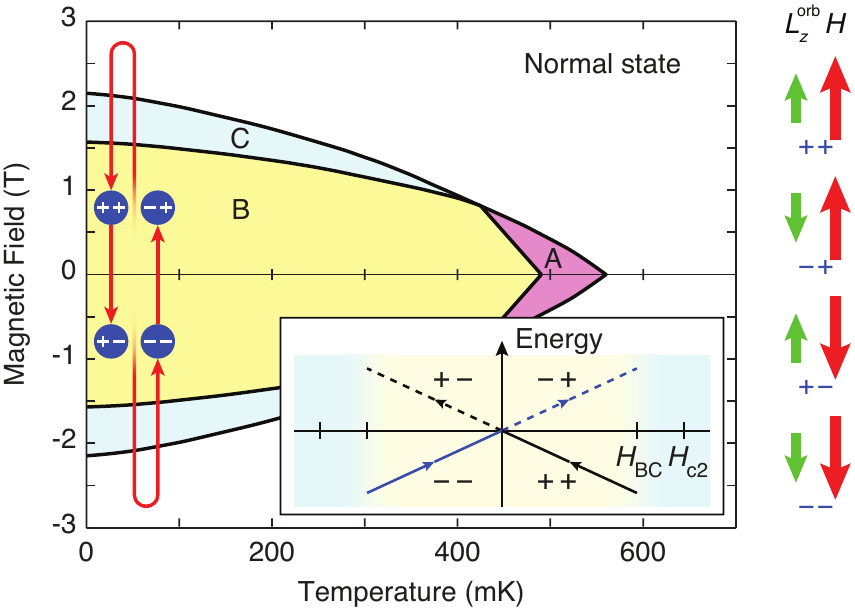}
\caption{\label{PD}
  (Color)
  {\UPt} phase diagram indicating the three vortex phases (A, B, C) for $\bf{H}\!\parallel\!\bf{c}$, adapted from Ref. ~\cite{Adenwalla:1990we}, with the zero-field transition temperatures, $T_{c} = 560$~mK, for the normal to A transition, and $T_{c_2} = 490$~mK for the A to B transition, and with $\Hcii(0)$ rescaled to reflect the high purity of the single crystal used for this work~\cite{Kycia:1998ui}.
  Measurements were performed on VLs prepared by a field reduction ({$++$}, {$--$}) or field reversal ({$-+$}, {$+-$}) procedure illustrated by the arrows and circles.
  Arrows on the right show the directions of the internal orbital angular momentum of the Cooper pairs ($L_z^{\text{orb}}$) and the phase winding of vortices as determined by the applied field ($H$).
  The inset shows the energy of the superconducting condensate as a function of field history: for a field reduction the system is in the ground state (solid lines) and for a field reversal in a metastable state (dashed lines).}
\end{figure}

In this report we focus on the heavy-fermion material {\UPt}~\cite{Joynt:2002wt}, which exhibits three distinct vortex phases shown in Fig.~\ref{PD} and labeled A, B and C, and two zero-field superconducting phases. The latter requires that the superconducting order parameter for {\UPt} belong to one of the odd-parity two-dimensional representations of the point group {\hex}~\cite{Hess:1989wt}, most likely a $f$-wave pairing states belonging to the $E_{2u}$ irreducible representation, in which the B phase is a chiral ground state that breaks time reversal and mirror symmetries~\cite{Sauls:1994wd}. The evidence for the latter comes from surface probes, most recently tunneling~\cite{Strand:2009eq} and polar Kerr rotation~\cite{Schemm:2014fv} in zero applied magnetic field, as well as early $\mu$SR measurements~\cite{Luke:1993fc}. However there are conflicting reports on the evidence for broken time-reversal symmetry in {\UPt} based on $\mu$SR~\cite{deReotier:1995fq}. Thus, it is important to have a direct test of broken time-reversal symmetry based on measurements in the bulk superconducting phases.

We report on small-angle neutron scattering (SANS) studies of the  vortex lattice (VL) structure in the B phase, extending into the C phase. Measurement were carried out with $\bf{H}\!\parallel\!\bf{c}$ where the VL is especially sensitive to changes in the superconducting state, due to the hexagonal crystal structure and corresponding weak basal plane anisotropy of {\UPt}~\cite{Shivaram:1986wo,Keller:1994do}.
We observe a difference in the VL diffraction patterns for $\bf{H}$ parallel versus anti-parallel to the original field direction on entering the B phase.
From this we infer that time-reversal symmetry is broken in the zero-field B phase, and that the $c$-axis is an axis of chiral symmetry.
A similar experiment using rotation in place of magnetic field was used to detect the chirality of the ground state of superfluid $^3$He-A~\cite{Walmsley:2012cp}.

\section{II. Background}
 The existence of two zero-field superconducting phases with a small splitting in the transition temperatures ($T_c, T_{c2}$) is the evidence for the pairing symmetry belonging to a two-dimensional $E$-representation and the presence of a weak symmetry breaking field, e.g. weak in-plane anti-ferromagnetism or strain, which lifts the degeneracy of the multi-dimensional representation leading to two zero-field phases and three VL phases~\cite{Hess:1989wt}.
Identification of the $E_{2u}$ representation proposed in Ref.~\cite{Sauls:1994wd}, which predicts a time-reversal symmetric A phase and broken time-reversal symmetry (BTRS) in the lower temperature B phase is supported by a a wide range of results. These include the $H$-$T$ phase diagram~\cite{Adenwalla:1990we,Sauls:1994wd,Shivaram:1986wo,Choi:1991wa}, thermodynamic and transport studies~\cite{Taillefer:1997eq,Graf:2000ua}, muon spin rotation experiments~\cite{Luke:1993fc}, phase-sensitive Josephson tunneling in the B-phase~\cite{Strand:2009eq}, and the previously mentioned directional tunneling to resolve the nodal structure in the A-phase~\cite{Strand:2010gz} as well as polar Kerr rotation that shows evidence for broken time-reversal symmetry in the B phase, but not the A phase~\cite{Schemm:2014fv}. Furthermore, the linear temperature dependence of the London penetration depth, obtained from both magnetization~\cite{Signore:1995hu,Schottl:1999ks} and SANS~\cite{Gannon:2015ct} measurements, is consistent with quadratic dispersion at the polar nodes of the energy gap structure that is one characteristic of the $E_{2u}$ model. With the exception of the SANS experiment by Gannon {\it et al.}~\cite{Gannon:2015ct}, all of the experiments listed above that support an odd-parity chiral state for the B phase were performed in the Meissner state~\cite{Luke:1993fc,Strand:2009eq,Strand:2010gz,Schemm:2014fv}.
The goal of our SANS studies is to investigate the theoretical prediction of BTRS in the B phase of {\UPt} using vortices as the probe.

\section{III. Methods}
The SANS experiments were performed at the GP-SANS beam line at the High Flux Isotope Reactor at Oak Ridge National Laboratory (ORNL) and at the D33 beam line at Institut Laue-Langevin (ILL)~\cite{5-42-402}. Preliminary, lower-resolution measurements were carried out at the SANS-I and SANS-II beam lines at the Paul Scherrer Institute.

Measurements were performed 
on VLs prepared using two different field histories illustrated in Fig.~\ref{PD}.
For a field reduction (labelled $++$ or $--$) the applied field was first increased into either the time-reversal symmetric C (ILL) or the normal (ORNL) phase, and then reduced to the measurement field ($\Hm$) without changing polarity.
For a field reversal ($+-$ or $-+$), the field was decreased into the Meissner state with chirality determined by the initial polarity of the field in the C or normal phase, then further decreased through zero such that $\Hm$ had the opposite polarity of the initial field in C phase or normal state.

Small-angle neutron scattering studies of the VL rely on the periodic field modulation due to vortices, with a scattered intensity $\propto \lambda^{-4}$~\cite{MuhlbauerRMP}.
For {\UPt} the large in-plane penetration depth $\lambda_{ab} \sim 680$~nm~\cite{Gannon:2015ct} 
necessitates a large sample volume.
We used the same 15~g, high-quality (RRR~$> 600$) {\UPt} single crystal as in previous experiments~\cite{Gannon:2015ct}.
A long, rod-like crystal was cut into two pieces which were co-aligned and fixed with silver epoxy (EPOTEK E4110) to a copper cold finger.
The sample assembly was mounted onto the mixing chamber of a dilution refrigerator and placed inside a superconducting magnet,
oriented with the crystalline $a$-axis vertical and the $c$-axis horizontally along the magnetic field and neutron beam.
Measurements were performed at base temperature $T \sim 50-65$~mK and with fields between $0.1$~T and 1.7~T.
To eliminate effects of eddy current heating, the field changes were paused at 0.1~T from the target field and the sample was allowed to thermalize. Once a stable sample temperature was achieved, the field was driven to the final value. Prior to each SANS measurement a damped field oscillation with an initial amplitude of 20~mT was applied to achieve a homogeneous vortex density and a well ordered VL. Periodically during the SANS measurements a 5~mT field oscillation was performed around the measurement field to maintain an ordered VL. All field oscillations finished by a reduction of the applied field magnitude, corresponding to a decrease of the vortex density.

All SANS measurements were carried out in a ``rocked on'' configuration, satisfying the Bragg condition for VL peaks at the top of the two-dimensional position sensitive 
detector~\cite{MuhlbauerRMP}.
Measurements of full rocking curves, where the VL peak intensities are recorded as they are being rotated through the Bragg condition, would greatly exceed the available SANS beam time and were therefore not feasible.
Background measurements, obtained either in zero field or above $\Hcii$, were subtracted from both the field reduction and field reversal data.

\section{IV. Results}
\begin{figure}
\includegraphics{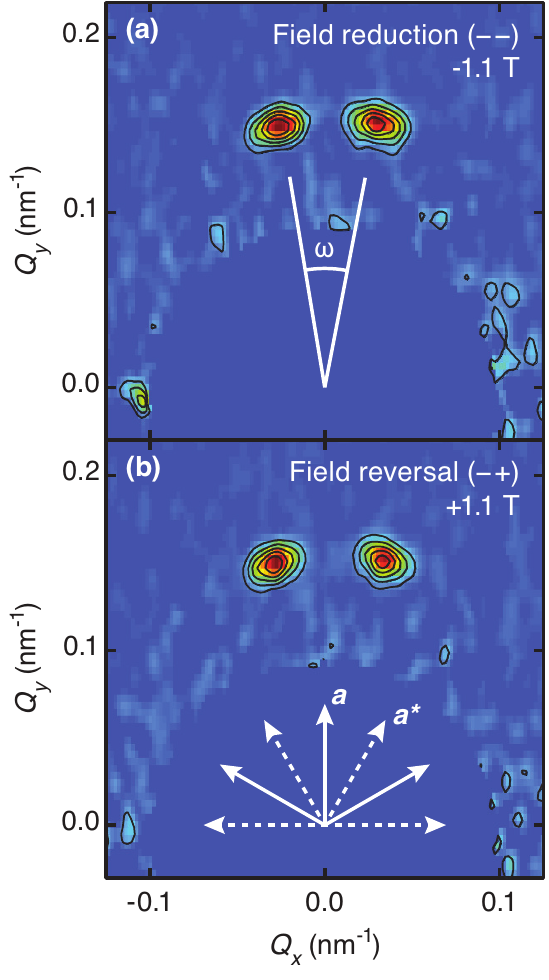}
\caption{\label{DifPat}
  (Color)
  Diffraction pattern obtained at $1.1$~T following a respectively (a) a field reduction and (b) a field reversal.
  The VL domain splitting ($\omega$) is shown in (a) and crystallographic directions within the scattering plane in (b).
  Measurements were performed at a single angular setting, satisfying the Bragg condition for VL reflections at the top of the detector.
  Zero field background scattering is subtracted, and the detector center near $Q = 0$ is masked off.}
\end{figure}

Figures~\ref{DifPat}(a) and (b) show SANS VL diffraction patterns obtained following a field reduction and reversal respectively, illustrating the two central results in this report.
First, the scattered intensity was equally divided between two Bragg peaks separated by a splitting angle, $\omega$, for both the field reduced and field reversed case.
This indicates the presence of two triangular VL domain orientations, rotated clockwise and counterclockwise about the crystalline $c$-axis. The splitting is also observed with just one of the two crystals illuminated by the neutron beam, indicating that the two VL domain orientations are intrinsic. In real space the splitting of the VL Bragg peaks corresponds to a nearest neighbor direction rotated away from the $\bf{a}^*$ direction by $\pm\omega/2$. Previous SANS studies with $\bf{H}\!\parallel\!\bf{c}$ were performed in a magnetic field, $0.19$~T, too low to resolve this splitting~\cite{Huxley:2000aa}.

\begin{figure}
\includegraphics{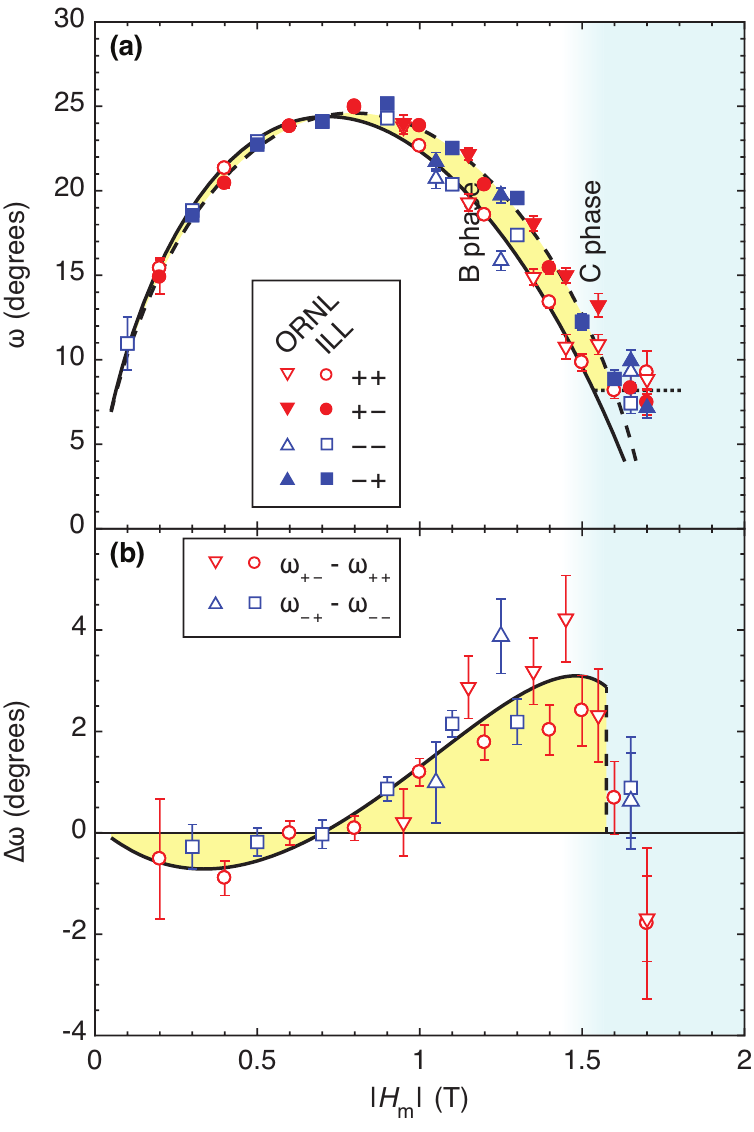}
\caption{\label{Split}
  (Color)
  Field dependence of (a) $\omega$ and (b) $\Delta \omega$, determined from the VL diffraction patterns.
  Lines in are fits to an empirical formula as described in the text.}
\end{figure}

Our main result is an observed difference of the Bragg peak splitting for the two field histories used, providing direct evidence for broken time-reversal symmetry of the parent B phase.
The field dependence of the VL splitting and the difference in the splitting of the domains ($\Delta\omega=\omega_{+-}-\omega_{++}$ or $\omega_{-+}-\omega_{--}$) are shown in Figs.~\ref{Split}(a) and (b) respectively.
The splitting was determined from the angular positions of two-Gaussian fits to the VL diffraction patterns, and error bars represent one standard deviation.
The curves in Fig.~\ref{Split} are fits to an empirical function of magnetic field~\cite{SM}.
Below $0.7$~T the field reversed split is slightly smaller than the field reduced one, giving rise to a negative $\Delta\omega$. For fields between $0.8$~T and $1.5$~T the situation is the opposite, where we find $\Delta\omega>0$.
While $\Delta\omega$ is small at low fields, the high field results are clear and statistically significant.
Finally, we note that there is no difference between measurements where the field reduced or field reversed state originated in the C phase or from the normal state, as is expected if both C and normal phases are time-reversal invariant.

These observations were made possible by extending the field range of previous SANS studies by more than a factor of eight for $\bf{H}\!\parallel\!\bf{c}$~\cite{Huxley:2000aa}, which we attribute to crystal quality and magnet homogeneity. This allowed us to explore the high field part of the phase diagram, where the VL is most sensitive to changes in the superconducting state and especially the vortex-core structure. The upper limit to accessible fields results from the reduction of the scattered intensity with increasing field, making SANS count times prohibitively long~\cite{SM}.
This is also the reason for the increased error bars for the determination of $\omega$ and $\Delta \omega$ at higher fields.

The location of the B-C phase transition varies between samples apparently owing to crystal quality~\cite{Adenwalla:1990we,Kycia:1998ui}. It is therefore desirable to identify features in the SANS data which allow a determination of this phase boundary. Previous SANS studies with $\bf{H}\!\perp\!\bf{c}$ found a change in the {\UPt} VL structure, and in one case a disordering upon entering the C phase~\cite{Yaron:1997wx,Gannon:2015ct}. 
For $\bf{H}\!\parallel\!\bf{c}$ we observe that the VL peak splitting changes with the applied field up to $1.5$~T, as shown in Fig.~\ref{Split}(a).
Above this field the splitting stabilizes at a $\omega \sim 8^{\circ}$ (dotted line), deviating from an extrapolation of the lower field data (full and dashed lines).

\begin{figure}
\includegraphics{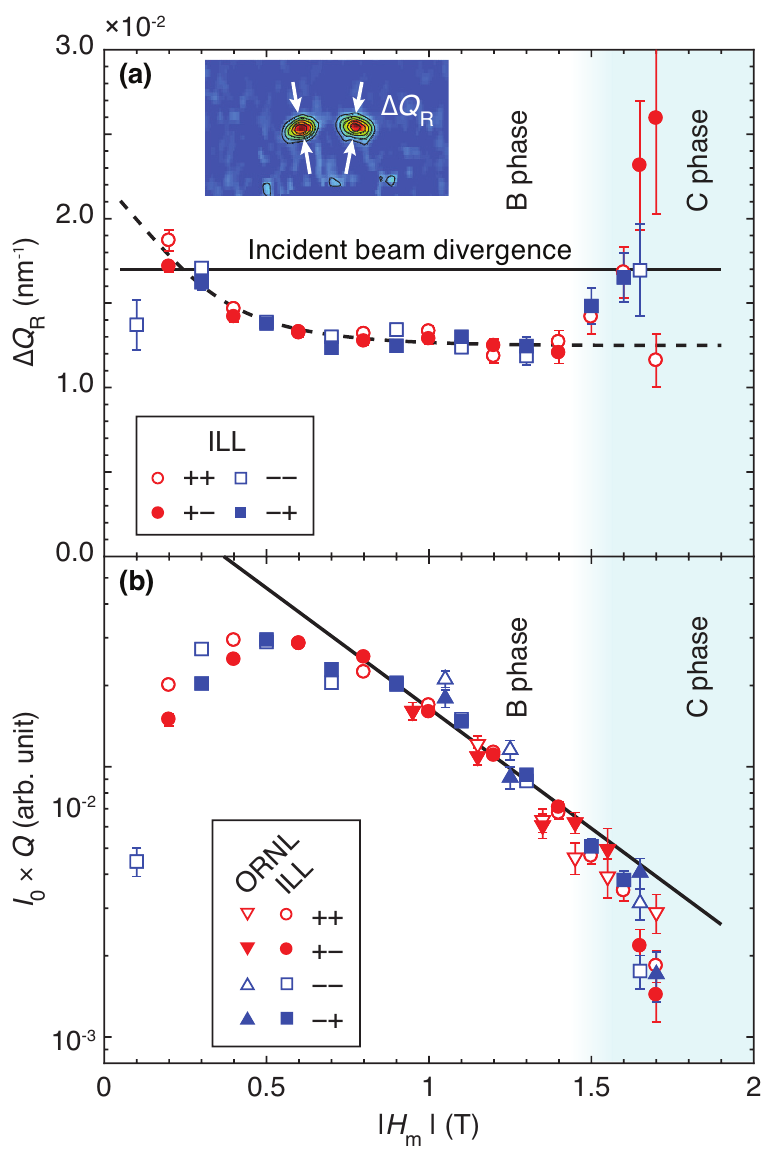}
\caption{\label{BCtrans}
  (Color)
  B-C phase transition determined from the onset of VL disordering. (a) Radial width of the Bragg peaks ($\Delta Q_R$).
  The solid line ($1.7 \times 10^{-2}$~nm$^{-1}$) corresponds to the incident beam divergence and the dashed line is a guide-to-the-eye.
  The insert indicates $\Delta Q_R$ in the detector plane. (b) Fit to $I_0 \times Q$ using an extended London model, as discussed in the text.
  The fit is restricted to fields in the range $0.8 - 1.5$~T and uses a fixed value of the penetration depth (680~nm).}
\end{figure}

The radial width of the Bragg peaks in the detector plane ($\Delta Q_{\text{R}}$), shown in Fig.~\ref{BCtrans}(a), also increases abruptly above $1.5$~T, indicating a disordering of the VL. Note that the minimum $\Delta Q_{\text{R}}$, observed at intermediate fields, is smaller than the divergence of the incident neutrons, determined from the undiffracted beam. This corresponds to a highly
ordered VL in the B phase, resulting in a diffracted beam that is better collimated than the incident one. The disordering at high fields is reflected in the peak scattered intensity ($I_0$), Fig.~\ref{BCtrans}(b).
Here the line is a fit of the intermediate field data to an extended London model~\cite{SM}.
Above $1.5$~T the measured intensity falls below the extrapolated fit, indicating a broadening of the Bragg peaks in the direction perpendicular to the detector plane (normal to $\Delta Q_{\text{R}}$).

We interpret the three features in the SANS data at $1.5$~T (splitting angle, radial width and intensity) as consistent signatures of the B-C transition. Note that disorder in the VL is also seen at low field in both $\Delta Q_{\text{R}}$ and the scattered intensity. This is due to a weakening of the vortex-vortex interactions at low vortex densities~\cite{Eskildsen:1997ab}, and is not associated with a change in the superconducting order parameter.

To ensure that the SANS results were not affected by a systematic asymmetry of the experimental setup, measurements were carried out using alternating directions of the C phase/normal state field, i.e. $++$/$+-$ and $--$/$-+$. As evident from Fig.~\ref{Split} this yielded consistent results, eliminating extrinsic effects as the cause for the difference between the splittings of the field reduced and field reversed protocols. Furthermore, results obtained using instruments at two different neutron scattering facilities (ORNL, ILL) are in excellent agreement.

Finally, the measurements are not affected by vortex pinning in any significant manner, evident from a comparison of the vortex density for the two different field histories.
This is obtained from the magnitude of the VL scattering vector $Q$, from which is it possible to determine the magnetic induction $B$. As shown in the Supplemental Material, there is perfect agreement between $B$ and the applied field, $H$, for both the field reduced and the field reversed cases, within measurement accuracy~\cite{SM}. This agreement also confirms the triangular symmetry of the VL made up of singly quantized vortices throughout the entire measured field range. Our data does not support theoretical predictions of doubly quantized vortices~\cite{Tokuyasu:1990tk,Sauls:2009dw,Ichioka:2012wy} based on {$E_{1u}$} pairing symmetry, or structural distortions/transitions within the B phase or associated with the B-C phase transition~\cite{Volovik:1988ul,Tsutsumi:2012hj,Takamatsu:2015cv} for $\bf{H} \! \parallel \! \bf{c}$. 

\section*{V. Discussion}
The structure of the order parameter in the vortex core depends sensitively on the symmetry of the parent superconducting state in the host material, making vortices an ideal probe of chiral superconductors. The VL symmetry and its orientation relative to the crystalline axes are determined by the anisotropy of the screening current in the plane perpendicular to the applied field, and the effect of this anisotropy on the vortex-vortex interactions. The anisotropy of the screening current can originate from several sources, particularly the Fermi velocity of the parent metallic state~\cite{Kogan:1997wy,Champel:2001kf}, as well as the order parameter~\cite{Volovik:1988ul,Hess:1989wt,Affleck:1997bg,Champel:2001kf}. 

For unconventional superconductors, which spontaneously break the symmetry of the Fermi surface, broken orbital rotational symmetry can lead to an anisotropy of the supercurrents that competes with or dominates the anisotropy arising from the Fermi velocities. This is the situation in {\UPt}. Here the dominant anisotropy in the A phase arises from the order parameter, which exhibits maximal orbital symmetry breaking with $|\Delta_{\mbox{\tiny A}}(\mathbf p)| = \Delta_{\mbox{\tiny A}}\,|2\hat{p}_x\hat{p}_y\hat{p}_z|$ for $E_{2u}$ pairing symmetry, and leads to the complex SANS diffraction patterns reported by Huxley {\it et al}.~{}\cite{Huxley:2000aa,Champel:2001kf}. However, even the B phase exhibits moderate in-plane gap anisotropy resulting from the in-plane symmetry breaking field, with $|\Delta_{\mbox{\tiny B}}(\mathbf p)|=\Delta_{\mbox{\tiny B}}(T)\,\left(1- \epsilon|\hat{p}_x^2 - \hat{p}_y^2|\right)|\hat{p}_z|$, where $\epsilon \propto \Delta T_c/T_c\approx 10\%$ is the magnitude of the symmetry breaking field measured in terms of the zero-field splitting of the superconducting transition, $\Delta T_c = T_c-T_{c_2}$~\cite{Sauls:1994wd,Strand:2010gz}. The anisotropy of the B phase gap is tetragonal, which generates a four-fold anisotropy of the screening current around an isolated vortex~\cite{SM}. Notably, the B phase gap anisotropy dominates the hexagonal anisotropy of the in-plane Fermi velocity, as the latter is of the order of $1\,\%$ based on measurements of the in-plane anisotropy of $H_{c_2}$~\cite{Shivaram:1986wo,Keller:1994do}. 

We make three important observations based on our experimental results.
The first is that in the low-field region the gap distortion of the parent B phase leads to the rotation
of the hexagonal VL which increases with field as the density of vortices increases. This provides a qualitative explanation for the increase in the domain splitting up to fields of order $H \approx 0.8\,\mbox{T}$.

The second observation is that the non-monotonic field dependence of the VL rotation shown in Fig.~\ref{Split}(a), onsetting at fields of order half the upper critical field, implies a change in the anisotropy of the vortex currents as the vortices become closely packed. This reflects a change in the anisotropy of the order parameter and current density near the vortex-cores.
Further evidence that the non-monotonic field dependence is due to the influence of the vortex-core structure is supported by the third observation.

There is asymmetry of the VL rotation for the two different field histories - field reduction compared to field reversal.
This  is the signature that the parent B phase breaks time reversal symmetry.
Specifically, for odd-parity superconductors that break time-reversal symmetry, vortices possess an internal degree of freedom associated with the phase winding of the vortex core, in addition to the global phase winding $n\in \{0, \pm 1, \pm 2, \ldots\}$.
The phase winding of the vortex core differs from the global phase winding by the topological winding number of the chiral B phase order parameter, $m=L_z^{\text{orb}}/\hbar$.
Thus, for a fixed ground state with chirality $\sgn(m)=+1$ there are two distinct VLs of singly quantized vortices labelled by $(\sgn(m)\,\sgn(n))$ corresponding to opposite phase windings embedded in the fixed chiral ground state. With $\sgn(m)=+1$: a VL of $(+\,+)$ vortices corresponds to $+{\mathbf H}\!\parallel\!L_z$ and a VL of $(+\,-)$ vortices corresponds to ${-\mathbf H}\!\parallel\!L_z$.
These time-reversed vortices, embedded in a fixed chiral ground state, have different energies and anisotropic vortex-core structures~\cite{Tokuyasu:1990ui}.
Since chirality is a discrete symmetry in {\UPt} it is not possible to deform the $\sgn(m)=+1$ chiral ground state into the time-reversed state with $\sgn(m)=-1$ by an adiabatic field reversal through the Meissner state.
This accounts for the metastable states shown in the Fig.~\ref{PD} insert.

The vortex-core order parameter is time-reversed relative to the parent B phase order parameter with winding number $p=n+2m$. Strong local anisotropy develops when a $p$-quantized vortex core dissociates, breaking the local rotational symmetry of the parent B phase~\cite{Tokuyasu:1990ui,Ichioka:2002go,Sauls:2009dw,Ichioka:2012wy}. The enhanced asymmetry in the VL rotation at fields of order $H \sim 1\,\mbox{T}$ suggests a broken rotational symmetry of at least one of the vortex cores in the chiral B phase. Quantitative calculations of the VL lattice structure, including the multi-band gap structure of {\UPt}~\cite{Nomoto:2016ko}, can provide a definitive answer to the source of enhanced chiral asymmetry at intermediate fields. Indeed quantitative calculations of the vortex core and current anisotropies in spin-triplet, $p$-wave superfluid $^3$He have been reported~\cite{Thuneberg:1986ix}.

\section{VI. Conclusion}
We have used SANS to study the VL in {\UPt} with $\bf{H} \! \parallel \! \bf{c}$ in both the B and C phases. We find a non-monotonic field dependence of the VL configuration, attributed to a competition between order parameter anisotropy and vortex-core structure. Depending on the relative direction of the chiral axis and the magnetic field we observe an asymmetry in the VL structure, which becomes more pronounced at higher vortex densities where vortex-core effects become more important. The results provide a direct signature of broken time-reversal symmetry of the bulk parent B phase, in agreement with theoretical predictions.

\section*{Acknowledgements}
This work was supported by the U.S. Department of Energy, Office of Basic Energy Sciences, under Awards No.~DE-SC0005051 (MRE: University of Notre Dame; neutron scattering) and DE-FG02-05ER46248 (WPH: Northwestern University; crystal growth and neutron scattering), and the Center for Applied Physics and Superconducting Technologies (JAS: Northwestern University; theory). The research of JAS is supported by National Science Foundation Grant DMR-1508730.
A portion of this research used resources at the High Flux Isotope Reactor, a U.S. DOE Office of Science User Facility operated by the Oak Ridge National Laboratory. Part of this work is based on experiments performed at the Institut Laue-Langevin, Grenoble, France and at the Swiss spallation neutron source SINQ, Paul Scherrer Institute, Villigen, Switzerland. We are grateful to the technical staff at all of these facilities for their assistance with the measurements.


\begin{thebibliography}{48}%
\makeatletter
\providecommand \@ifxundefined [1]{%
 \@ifx{#1\undefined}
}%
\providecommand \@ifnum [1]{%
 \ifnum #1\expandafter \@firstoftwo
 \else \expandafter \@secondoftwo
 \fi
}%
\providecommand \@ifx [1]{%
 \ifx #1\expandafter \@firstoftwo
 \else \expandafter \@secondoftwo
 \fi
}%
\providecommand \natexlab [1]{#1}%
\providecommand \enquote  [1]{``#1''}%
\providecommand \bibnamefont  [1]{#1}%
\providecommand \bibfnamefont [1]{#1}%
\providecommand \citenamefont [1]{#1}%
\providecommand \href@noop [0]{\@secondoftwo}%
\providecommand \href [0]{\begingroup \@sanitize@url \@href}%
\providecommand \@href[1]{\@@startlink{#1}\@@href}%
\providecommand \@@href[1]{\endgroup#1\@@endlink}%
\providecommand \@sanitize@url [0]{\catcode `\\12\catcode `\$12\catcode
  `\&12\catcode `\#12\catcode `\^12\catcode `\_12\catcode `\%12\relax}%
\providecommand \@@startlink[1]{}%
\providecommand \@@endlink[0]{}%
\providecommand \url  [0]{\begingroup\@sanitize@url \@url }%
\providecommand \@url [1]{\endgroup\@href {#1}{\urlprefix }}%
\providecommand \urlprefix  [0]{URL }%
\providecommand \Eprint [0]{\href }%
\providecommand \doibase [0]{http://dx.doi.org/}%
\providecommand \selectlanguage [0]{\@gobble}%
\providecommand \bibinfo  [0]{\@secondoftwo}%
\providecommand \bibfield  [0]{\@secondoftwo}%
\providecommand \translation [1]{[#1]}%
\providecommand \BibitemOpen [0]{}%
\providecommand \bibitemStop [0]{}%
\providecommand \bibitemNoStop [0]{.\EOS\space}%
\providecommand \EOS [0]{\spacefactor3000\relax}%
\providecommand \BibitemShut  [1]{\csname bibitem#1\endcsname}%
\let\auto@bib@innerbib\@empty
\bibitem [{\citenamefont {Qi}\ and\ \citenamefont {Zhang}(2011)}]{Qi:2011hba}%
  \BibitemOpen
  \bibfield  {author} {\bibinfo {author} {\bibfnamefont {X-L}\ \bibnamefont
  {Qi}}\ and\ \bibinfo {author} {\bibfnamefont {S-C}\ \bibnamefont {Zhang}},\
  }\bibfield  {title} {\enquote {\bibinfo {title} {{Topological insulators and
  superconductors}},}\ }\href@noop {} {\bibfield  {journal} {\bibinfo
  {journal} {Rev. Mod. Phys.}\ }\textbf {\bibinfo {volume} {83}},\ \bibinfo
  {pages} {1057--1110} (\bibinfo {year} {2011})}\BibitemShut {NoStop}%
\bibitem [{\citenamefont {Schnyder}\ and\ \citenamefont
  {Brydon}(2015)}]{Schnyder:2015kt}%
  \BibitemOpen
  \bibfield  {author} {\bibinfo {author} {\bibfnamefont {A~P}\ \bibnamefont
  {Schnyder}}\ and\ \bibinfo {author} {\bibfnamefont {P~M~R}\ \bibnamefont
  {Brydon}},\ }\bibfield  {title} {\enquote {\bibinfo {title} {{Topological
  surface states in nodal superconductors}},}\ }\href@noop {} {\bibfield
  {journal} {\bibinfo  {journal} {J. Phys.: Condens. Matter}\ }\textbf
  {\bibinfo {volume} {27}},\ \bibinfo {pages} {243201} (\bibinfo {year}
  {2015})}\BibitemShut {NoStop}%
\bibitem [{\citenamefont {Beenakker}\ and\ \citenamefont
  {Kouwenhoven}(2016)}]{Beenakker:2016ds}%
  \BibitemOpen
  \bibfield  {author} {\bibinfo {author} {\bibfnamefont {C}~\bibnamefont
  {Beenakker}}\ and\ \bibinfo {author} {\bibfnamefont {L}~\bibnamefont
  {Kouwenhoven}},\ }\bibfield  {title} {\enquote {\bibinfo {title} {{A road to
  reality with topological superconductors}},}\ }\href@noop {} {\bibfield
  {journal} {\bibinfo  {journal} {Nat. Phys.}\ }\textbf {\bibinfo {volume}
  {12}},\ \bibinfo {pages} {618--621} (\bibinfo {year} {2016})}\BibitemShut
  {NoStop}%
\bibitem [{\citenamefont {Kozii}\ \emph {et~al.}(2016)\citenamefont {Kozii},
  \citenamefont {Venderbos},\ and\ \citenamefont {Fu}}]{Kozii:2016go}%
  \BibitemOpen
  \bibfield  {author} {\bibinfo {author} {\bibfnamefont {V}~\bibnamefont
  {Kozii}}, \bibinfo {author} {\bibfnamefont {J~W~F}\ \bibnamefont
  {Venderbos}}, \ and\ \bibinfo {author} {\bibfnamefont {L}~\bibnamefont
  {Fu}},\ }\bibfield  {title} {\enquote {\bibinfo {title} {{Three-dimensional
  Majorana fermions in chiral superconductors}},}\ }\href@noop {} {\bibfield
  {journal} {\bibinfo  {journal} {Sci. Adv.}\ }\textbf {\bibinfo {volume}
  {2}},\ \bibinfo {pages} {e1601835} (\bibinfo {year} {2016})}\BibitemShut
  {NoStop}%
\bibitem [{\citenamefont {Mizushima}\ \emph {et~al.}(2016)\citenamefont
  {Mizushima}, \citenamefont {Tsutsumi}, \citenamefont {Kawakami},
  \citenamefont {Sato}, \citenamefont {Ichioka},\ and\ \citenamefont
  {Machida}}]{Mizushima:2016hk}%
  \BibitemOpen
  \bibfield  {author} {\bibinfo {author} {\bibfnamefont {T}~\bibnamefont
  {Mizushima}}, \bibinfo {author} {\bibfnamefont {Y}~\bibnamefont {Tsutsumi}},
  \bibinfo {author} {\bibfnamefont {T}~\bibnamefont {Kawakami}}, \bibinfo
  {author} {\bibfnamefont {M}~\bibnamefont {Sato}}, \bibinfo {author}
  {\bibfnamefont {M}~\bibnamefont {Ichioka}}, \ and\ \bibinfo {author}
  {\bibfnamefont {K}~\bibnamefont {Machida}},\ }\bibfield  {title} {\enquote
  {\bibinfo {title} {{Symmetry-Protected Topological Superfluids and
  Superconductors -- From the Basics to $^3$He --}},}\ }\href@noop {}
  {\bibfield  {journal} {\bibinfo  {journal} {J. Phys. Soc. Jpn.}\ }\textbf
  {\bibinfo {volume} {85}},\ \bibinfo {pages} {022001} (\bibinfo {year}
  {2016})}\BibitemShut {NoStop}%
\bibitem [{\citenamefont {Sato}\ and\ \citenamefont
  {Ando}(2017)}]{Sato:2017th}%
  \BibitemOpen
  \bibfield  {author} {\bibinfo {author} {\bibfnamefont {M}~\bibnamefont
  {Sato}}\ and\ \bibinfo {author} {\bibfnamefont {Y}~\bibnamefont {Ando}},\
  }\bibfield  {title} {\enquote {\bibinfo {title} {{Topological
  superconductors: a review}},}\ }\href@noop {} {\bibfield  {journal} {\bibinfo
   {journal} {Rep. Prog. Phys.}\ }\textbf {\bibinfo {volume} {80}},\ \bibinfo
  {pages} {076501} (\bibinfo {year} {2017})}\BibitemShut {NoStop}%
\bibitem [{\citenamefont {Strand}\ \emph {et~al.}(2009)\citenamefont {Strand},
  \citenamefont {Van~Harlingen}, \citenamefont {Kycia},\ and\ \citenamefont
  {Halperin}}]{Strand:2009eq}%
  \BibitemOpen
  \bibfield  {author} {\bibinfo {author} {\bibfnamefont {J~D}\ \bibnamefont
  {Strand}}, \bibinfo {author} {\bibfnamefont {D~J}\ \bibnamefont
  {Van~Harlingen}}, \bibinfo {author} {\bibfnamefont {J}~\bibnamefont {Kycia}},
  \ and\ \bibinfo {author} {\bibfnamefont {W~P}\ \bibnamefont {Halperin}},\
  }\bibfield  {title} {\enquote {\bibinfo {title} {{Evidence for Complex
  Superconducting Order Parameter Symmetry in the Low-Temperature Phase of
  UPt$_3$ from Josephson Interferometry}},}\ }\href@noop {} {\bibfield
  {journal} {\bibinfo  {journal} {Phys. Rev. Lett.}\ }\textbf {\bibinfo
  {volume} {103}},\ \bibinfo {pages} {197002} (\bibinfo {year}
  {2009})}\BibitemShut {NoStop}%
\bibitem [{\citenamefont {Schemm}\ \emph {et~al.}(2014)\citenamefont {Schemm},
  \citenamefont {Gannon}, \citenamefont {Wishne}, \citenamefont {Halperin},\
  and\ \citenamefont {Kapitulnik}}]{Schemm:2014fv}%
  \BibitemOpen
  \bibfield  {author} {\bibinfo {author} {\bibfnamefont {E~R}\ \bibnamefont
  {Schemm}}, \bibinfo {author} {\bibfnamefont {W~J}\ \bibnamefont {Gannon}},
  \bibinfo {author} {\bibfnamefont {C~M}\ \bibnamefont {Wishne}}, \bibinfo
  {author} {\bibfnamefont {W~P}\ \bibnamefont {Halperin}}, \ and\ \bibinfo
  {author} {\bibfnamefont {A}~\bibnamefont {Kapitulnik}},\ }\bibfield  {title}
  {\enquote {\bibinfo {title} {{Observation of broken time-reversal symmetry in
  the heavy-fermion superconductor UPt$_3$}},}\ }\href@noop {} {\bibfield
  {journal} {\bibinfo  {journal} {Science}\ }\textbf {\bibinfo {volume}
  {345}},\ \bibinfo {pages} {190--193} (\bibinfo {year} {2014})}\BibitemShut
  {NoStop}%
\bibitem [{\citenamefont {Maeno}\ \emph {et~al.}(2012)\citenamefont {Maeno},
  \citenamefont {Kittaka}, \citenamefont {Nomura}, \citenamefont {Yonezawa},\
  and\ \citenamefont {Ishida}}]{Maeno:2012ew}%
  \BibitemOpen
  \bibfield  {author} {\bibinfo {author} {\bibfnamefont {Y}~\bibnamefont
  {Maeno}}, \bibinfo {author} {\bibfnamefont {S}~\bibnamefont {Kittaka}},
  \bibinfo {author} {\bibfnamefont {T}~\bibnamefont {Nomura}}, \bibinfo
  {author} {\bibfnamefont {S}~\bibnamefont {Yonezawa}}, \ and\ \bibinfo
  {author} {\bibfnamefont {K}~\bibnamefont {Ishida}},\ }\bibfield  {title}
  {\enquote {\bibinfo {title} {{Evaluation of Spin-Triplet Superconductivity in
  Sr$_2$RuO$_4$}},}\ }\href@noop {} {\bibfield  {journal} {\bibinfo  {journal}
  {J. Phys. Soc. Jpn.}\ }\textbf {\bibinfo {volume} {81}},\ \bibinfo {pages}
  {011009} (\bibinfo {year} {2012})}\BibitemShut {NoStop}%
\bibitem [{\citenamefont {Steppke}\ \emph {et~al.}(2017)\citenamefont
  {Steppke}, \citenamefont {Zhao}, \citenamefont {Barber}, \citenamefont
  {Scaffidi}, \citenamefont {Jerzembeck}, \citenamefont {Rosner}, \citenamefont
  {Gibbs}, \citenamefont {Maeno}, \citenamefont {Simon}, \citenamefont
  {Mackenzie},\ and\ \citenamefont {Hicks}}]{Steppke:2017fb}%
  \BibitemOpen
  \bibfield  {author} {\bibinfo {author} {\bibfnamefont {A}~\bibnamefont
  {Steppke}}, \bibinfo {author} {\bibfnamefont {L}~\bibnamefont {Zhao}},
  \bibinfo {author} {\bibfnamefont {M~E}\ \bibnamefont {Barber}}, \bibinfo
  {author} {\bibfnamefont {T}~\bibnamefont {Scaffidi}}, \bibinfo {author}
  {\bibfnamefont {F}~\bibnamefont {Jerzembeck}}, \bibinfo {author}
  {\bibfnamefont {H}~\bibnamefont {Rosner}}, \bibinfo {author} {\bibfnamefont
  {A~S}\ \bibnamefont {Gibbs}}, \bibinfo {author} {\bibfnamefont
  {Y}~\bibnamefont {Maeno}}, \bibinfo {author} {\bibfnamefont {S~H}\
  \bibnamefont {Simon}}, \bibinfo {author} {\bibfnamefont {A~P}\ \bibnamefont
  {Mackenzie}}, \ and\ \bibinfo {author} {\bibfnamefont {C~W}\ \bibnamefont
  {Hicks}},\ }\bibfield  {title} {\enquote {\bibinfo {title} {{Strong peak in
  $T_c$ of Sr$_2$RuO$_4$ under uniaxial pressure}},}\ }\href@noop {} {\bibfield
   {journal} {\bibinfo  {journal} {Science}\ }\textbf {\bibinfo {volume}
  {355}},\ \bibinfo {pages} {eaaf9398} (\bibinfo {year} {2017})}\BibitemShut
  {NoStop}%
\bibitem [{\citenamefont {Ikegami}\ \emph {et~al.}(2013)\citenamefont
  {Ikegami}, \citenamefont {Tsutsumi},\ and\ \citenamefont
  {Kono}}]{Ikegami:2013jx}%
  \BibitemOpen
  \bibfield  {author} {\bibinfo {author} {\bibfnamefont {H}~\bibnamefont
  {Ikegami}}, \bibinfo {author} {\bibfnamefont {Y}~\bibnamefont {Tsutsumi}}, \
  and\ \bibinfo {author} {\bibfnamefont {K}~\bibnamefont {Kono}},\ }\bibfield
  {title} {\enquote {\bibinfo {title} {{Chiral Symmetry Breaking in Superfluid
  $^3$He-A}},}\ }\href@noop {} {\bibfield  {journal} {\bibinfo  {journal}
  {Science}\ }\textbf {\bibinfo {volume} {341}},\ \bibinfo {pages} {59--62}
  (\bibinfo {year} {2013})}\BibitemShut {NoStop}%
\bibitem [{\citenamefont {Autti}\ \emph {et~al.}(2016)\citenamefont {Autti},
  \citenamefont {Dmitriev}, \citenamefont {M{\"a}kinen}, \citenamefont
  {Soldatov}, \citenamefont {Volovik}, \citenamefont {Yudin}, \citenamefont
  {Zavjalov},\ and\ \citenamefont {Eltsov}}]{Autti:2016bu}%
  \BibitemOpen
  \bibfield  {author} {\bibinfo {author} {\bibfnamefont {S}~\bibnamefont
  {Autti}}, \bibinfo {author} {\bibfnamefont {V~V}\ \bibnamefont {Dmitriev}},
  \bibinfo {author} {\bibfnamefont {J~T}\ \bibnamefont {M{\"a}kinen}}, \bibinfo
  {author} {\bibfnamefont {A~A}\ \bibnamefont {Soldatov}}, \bibinfo {author}
  {\bibfnamefont {G~E}\ \bibnamefont {Volovik}}, \bibinfo {author}
  {\bibfnamefont {A~N}\ \bibnamefont {Yudin}}, \bibinfo {author} {\bibfnamefont
  {V~V}\ \bibnamefont {Zavjalov}}, \ and\ \bibinfo {author} {\bibfnamefont
  {V~B}\ \bibnamefont {Eltsov}},\ }\bibfield  {title} {\enquote {\bibinfo
  {title} {{Observation of Half-Quantum Vortices in Topological Superfluid
  $^3$He}},}\ }\href@noop {} {\bibfield  {journal} {\bibinfo  {journal} {Phys.
  Rev. Lett.}\ }\textbf {\bibinfo {volume} {117}},\ \bibinfo {pages} {255301}
  (\bibinfo {year} {2016})}\BibitemShut {NoStop}%
\bibitem [{\citenamefont {Joynt}\ and\ \citenamefont
  {Taillefer}(2002)}]{Joynt:2002wt}%
  \BibitemOpen
  \bibfield  {author} {\bibinfo {author} {\bibfnamefont {R}~\bibnamefont
  {Joynt}}\ and\ \bibinfo {author} {\bibfnamefont {L}~\bibnamefont
  {Taillefer}},\ }\bibfield  {title} {\enquote {\bibinfo {title} {{The
  superconducting phases of UPt$_3$}},}\ }\href@noop {} {\bibfield  {journal}
  {\bibinfo  {journal} {Rev. Mod. Phys.}\ }\textbf {\bibinfo {volume} {74}},\
  \bibinfo {pages} {236} (\bibinfo {year} {2002})}\BibitemShut {NoStop}%
\bibitem [{\citenamefont {Hess}\ \emph {et~al.}(1989)\citenamefont {Hess},
  \citenamefont {Tokuyasu},\ and\ \citenamefont {Sauls}}]{Hess:1989wt}%
  \BibitemOpen
  \bibfield  {author} {\bibinfo {author} {\bibfnamefont {D~W}\ \bibnamefont
  {Hess}}, \bibinfo {author} {\bibfnamefont {T~A}\ \bibnamefont {Tokuyasu}}, \
  and\ \bibinfo {author} {\bibfnamefont {J~A}\ \bibnamefont {Sauls}},\
  }\bibfield  {title} {\enquote {\bibinfo {title} {{Broken symmetry in an
  unconventional superconductor: a model for the double transition in
  UPt$_3$}},}\ }\href@noop {} {\bibfield  {journal} {\bibinfo  {journal} {J.
  Phys.: Condens. Matter}\ }\textbf {\bibinfo {volume} {1}},\ \bibinfo {pages}
  {8135--8145} (\bibinfo {year} {1989})}\BibitemShut {NoStop}%
\bibitem [{\citenamefont {Sauls}(1994)}]{Sauls:1994wd}%
  \BibitemOpen
  \bibfield  {author} {\bibinfo {author} {\bibfnamefont {J~A}\ \bibnamefont
  {Sauls}},\ }\bibfield  {title} {\enquote {\bibinfo {title} {{The Order
  Parameter for the Superconducting Phases of UPt$_3$}},}\ }\href@noop {}
  {\bibfield  {journal} {\bibinfo  {journal} {Adv. Phys.}\ }\textbf {\bibinfo
  {volume} {43}},\ \bibinfo {pages} {113} (\bibinfo {year} {1994})}\BibitemShut
  {NoStop}%
\bibitem [{\citenamefont {Luke}\ \emph {et~al.}(1993)\citenamefont {Luke},
  \citenamefont {Keren}, \citenamefont {Le}, \citenamefont {Wu}, \citenamefont
  {Uemura}, \citenamefont {Bonn}, \citenamefont {Taillefer},\ and\
  \citenamefont {Garrett}}]{Luke:1993fc}%
  \BibitemOpen
  \bibfield  {author} {\bibinfo {author} {\bibfnamefont {G~M}\ \bibnamefont
  {Luke}}, \bibinfo {author} {\bibfnamefont {A}~\bibnamefont {Keren}}, \bibinfo
  {author} {\bibfnamefont {L~P}\ \bibnamefont {Le}}, \bibinfo {author}
  {\bibfnamefont {W~D}\ \bibnamefont {Wu}}, \bibinfo {author} {\bibfnamefont
  {Y~J}\ \bibnamefont {Uemura}}, \bibinfo {author} {\bibfnamefont {D~A}\
  \bibnamefont {Bonn}}, \bibinfo {author} {\bibfnamefont {Louis}\ \bibnamefont
  {Taillefer}}, \ and\ \bibinfo {author} {\bibfnamefont {J~D}\ \bibnamefont
  {Garrett}},\ }\bibfield  {title} {\enquote {\bibinfo {title} {{Muon spin
  relaxation in UPt$_{3}$}},}\ }\href@noop {} {\bibfield  {journal} {\bibinfo
  {journal} {Phys. Rev. Lett.}\ }\textbf {\bibinfo {volume} {71}},\ \bibinfo
  {pages} {1466--1469} (\bibinfo {year} {1993})}\BibitemShut {NoStop}%
\bibitem [{\citenamefont {de~R{\'e}otier}\ \emph {et~al.}(1995)\citenamefont
  {de~R{\'e}otier}, \citenamefont {Huxley}, \citenamefont {Yaouanc},
  \citenamefont {Flouquet}, \citenamefont {Bonville}, \citenamefont {Imbert},
  \citenamefont {Pari}, \citenamefont {Gubbens},\ and\ \citenamefont
  {Mulders}}]{deReotier:1995fq}%
  \BibitemOpen
  \bibfield  {author} {\bibinfo {author} {\bibfnamefont {P~D}\ \bibnamefont
  {de~R{\'e}otier}}, \bibinfo {author} {\bibfnamefont {A}~\bibnamefont
  {Huxley}}, \bibinfo {author} {\bibfnamefont {A}~\bibnamefont {Yaouanc}},
  \bibinfo {author} {\bibfnamefont {J}~\bibnamefont {Flouquet}}, \bibinfo
  {author} {\bibfnamefont {P}~\bibnamefont {Bonville}}, \bibinfo {author}
  {\bibfnamefont {P}~\bibnamefont {Imbert}}, \bibinfo {author} {\bibfnamefont
  {P}~\bibnamefont {Pari}}, \bibinfo {author} {\bibfnamefont {P~C~M}\
  \bibnamefont {Gubbens}}, \ and\ \bibinfo {author} {\bibfnamefont {A~M}\
  \bibnamefont {Mulders}},\ }\bibfield  {title} {\enquote {\bibinfo {title}
  {{Absence of zero field muon spin relaxation induced by superconductivity in
  the B phase of UPt$_3$}},}\ }\href@noop {} {\bibfield  {journal} {\bibinfo
  {journal} {Phys. Lett. A}\ }\textbf {\bibinfo {volume} {205}},\ \bibinfo
  {pages} {239--243} (\bibinfo {year} {1995})}\BibitemShut {NoStop}%
\bibitem [{\citenamefont {Shivaram}\ \emph {et~al.}(1986)\citenamefont
  {Shivaram}, \citenamefont {Rosenbaum},\ and\ \citenamefont
  {Hinks}}]{Shivaram:1986wo}%
  \BibitemOpen
  \bibfield  {author} {\bibinfo {author} {\bibfnamefont {B~S}\ \bibnamefont
  {Shivaram}}, \bibinfo {author} {\bibfnamefont {T~F}\ \bibnamefont
  {Rosenbaum}}, \ and\ \bibinfo {author} {\bibfnamefont {D~G}\ \bibnamefont
  {Hinks}},\ }\bibfield  {title} {\enquote {\bibinfo {title} {{Unusual Angular
  and Temperature Dependence of the Upper Critical Field in UPt$_3$}},}\
  }\href@noop {} {\bibfield  {journal} {\bibinfo  {journal} {Phys. Rev. Lett.}\
  }\textbf {\bibinfo {volume} {57}},\ \bibinfo {pages} {1259--1262} (\bibinfo
  {year} {1986})}\BibitemShut {NoStop}%
\bibitem [{\citenamefont {Keller}\ \emph {et~al.}(1994)\citenamefont {Keller},
  \citenamefont {Tholence}, \citenamefont {Huxley},\ and\ \citenamefont
  {Flouquet}}]{Keller:1994do}%
  \BibitemOpen
  \bibfield  {author} {\bibinfo {author} {\bibfnamefont {N}~\bibnamefont
  {Keller}}, \bibinfo {author} {\bibfnamefont {J~L}\ \bibnamefont {Tholence}},
  \bibinfo {author} {\bibfnamefont {A}~\bibnamefont {Huxley}}, \ and\ \bibinfo
  {author} {\bibfnamefont {J}~\bibnamefont {Flouquet}},\ }\bibfield  {title}
  {\enquote {\bibinfo {title} {{Angular Dependence of the Upper Critical Field
  of the Heavy Fermion Superconductor UPt$_3$}},}\ }\href@noop {} {\bibfield
  {journal} {\bibinfo  {journal} {Phys. Rev. Lett.}\ }\textbf {\bibinfo
  {volume} {73}},\ \bibinfo {pages} {2364--2367} (\bibinfo {year}
  {1994})}\BibitemShut {NoStop}%
\bibitem [{\citenamefont {Walmsley}\ and\ \citenamefont
  {Golov}(2012)}]{Walmsley:2012cp}%
  \BibitemOpen
  \bibfield  {author} {\bibinfo {author} {\bibfnamefont {P~M}\ \bibnamefont
  {Walmsley}}\ and\ \bibinfo {author} {\bibfnamefont {A~I}\ \bibnamefont
  {Golov}},\ }\bibfield  {title} {\enquote {\bibinfo {title} {{Chirality of
  Superfluid $^3$He-$A$}},}\ }\href@noop {} {\bibfield  {journal} {\bibinfo
  {journal} {Phys. Rev. Lett.}\ }\textbf {\bibinfo {volume} {109}},\ \bibinfo
  {pages} {215301} (\bibinfo {year} {2012})}\BibitemShut {NoStop}%
\bibitem [{\citenamefont {Adenwalla}\ \emph {et~al.}(1990)\citenamefont
  {Adenwalla}, \citenamefont {Lin}, \citenamefont {Ran}, \citenamefont {Zhao},
  \citenamefont {Ketterson}, \citenamefont {Sauls}, \citenamefont {Taillefer},
  \citenamefont {Hinks}, \citenamefont {Levy},\ and\ \citenamefont
  {Sarma}}]{Adenwalla:1990we}%
  \BibitemOpen
  \bibfield  {author} {\bibinfo {author} {\bibfnamefont {S}~\bibnamefont
  {Adenwalla}}, \bibinfo {author} {\bibfnamefont {S~W}\ \bibnamefont {Lin}},
  \bibinfo {author} {\bibfnamefont {Q~Z}\ \bibnamefont {Ran}}, \bibinfo
  {author} {\bibfnamefont {Z}~\bibnamefont {Zhao}}, \bibinfo {author}
  {\bibfnamefont {J~B}\ \bibnamefont {Ketterson}}, \bibinfo {author}
  {\bibfnamefont {J~A}\ \bibnamefont {Sauls}}, \bibinfo {author} {\bibfnamefont
  {L}~\bibnamefont {Taillefer}}, \bibinfo {author} {\bibfnamefont {D~G}\
  \bibnamefont {Hinks}}, \bibinfo {author} {\bibfnamefont {M}~\bibnamefont
  {Levy}}, \ and\ \bibinfo {author} {\bibfnamefont {Bimal~K}\ \bibnamefont
  {Sarma}},\ }\bibfield  {title} {\enquote {\bibinfo {title} {{Phase Diagram of
  UPt$_3$ from Ultrasonic Velocity Measurements}},}\ }\href@noop {} {\bibfield
  {journal} {\bibinfo  {journal} {Phys. Rev. Lett.}\ }\textbf {\bibinfo
  {volume} {65}},\ \bibinfo {pages} {2298--2301} (\bibinfo {year}
  {1990})}\BibitemShut {NoStop}%
\bibitem [{\citenamefont {Choi}\ and\ \citenamefont
  {Sauls}(1991)}]{Choi:1991wa}%
  \BibitemOpen
  \bibfield  {author} {\bibinfo {author} {\bibfnamefont {C~H}\ \bibnamefont
  {Choi}}\ and\ \bibinfo {author} {\bibfnamefont {J~A}\ \bibnamefont {Sauls}},\
  }\bibfield  {title} {\enquote {\bibinfo {title} {{Identification of
  Odd-Parity Superconductivity in UPt$_3$ from Paramagnetic Effects on the
  Upper Critical Field}},}\ }\href@noop {} {\bibfield  {journal} {\bibinfo
  {journal} {Phys. Rev. Lett.}\ }\textbf {\bibinfo {volume} {66}},\ \bibinfo
  {pages} {484--487} (\bibinfo {year} {1991})}\BibitemShut {NoStop}%
\bibitem [{\citenamefont {Taillefer}\ \emph {et~al.}(1997)\citenamefont
  {Taillefer}, \citenamefont {Ellman}, \citenamefont {Lussier},\ and\
  \citenamefont {Poirier}}]{Taillefer:1997eq}%
  \BibitemOpen
  \bibfield  {author} {\bibinfo {author} {\bibfnamefont {L}~\bibnamefont
  {Taillefer}}, \bibinfo {author} {\bibfnamefont {B}~\bibnamefont {Ellman}},
  \bibinfo {author} {\bibfnamefont {B}~\bibnamefont {Lussier}}, \ and\ \bibinfo
  {author} {\bibfnamefont {M}~\bibnamefont {Poirier}},\ }\bibfield  {title}
  {\enquote {\bibinfo {title} {{On the gap structure of UPt$_3$: phases A and
  B}},}\ }\href@noop {} {\bibfield  {journal} {\bibinfo  {journal} {Physica B:
  Condensed Matter}\ }\textbf {\bibinfo {volume} {230--232}},\ \bibinfo {pages}
  {327--331} (\bibinfo {year} {1997})}\BibitemShut {NoStop}%
\bibitem [{\citenamefont {Graf}\ \emph {et~al.}(2000)\citenamefont {Graf},
  \citenamefont {Yip},\ and\ \citenamefont {Sauls}}]{Graf:2000ua}%
  \BibitemOpen
  \bibfield  {author} {\bibinfo {author} {\bibfnamefont {M~J}\ \bibnamefont
  {Graf}}, \bibinfo {author} {\bibfnamefont {S~K}\ \bibnamefont {Yip}}, \ and\
  \bibinfo {author} {\bibfnamefont {J~A}\ \bibnamefont {Sauls}},\ }\bibfield
  {title} {\enquote {\bibinfo {title} {{Identification of the orbital pairing
  symmetry in UPt$_3$}},}\ }\href@noop {} {\bibfield  {journal} {\bibinfo
  {journal} {Phys. Rev. B}\ }\textbf {\bibinfo {volume} {62}},\ \bibinfo
  {pages} {14393--14402} (\bibinfo {year} {2000})}\BibitemShut {NoStop}%
\bibitem [{\citenamefont {Strand}\ \emph {et~al.}(2010)\citenamefont {Strand},
  \citenamefont {Bahr}, \citenamefont {Van~Harlingen}, \citenamefont {Davis},
  \citenamefont {Gannon},\ and\ \citenamefont {Halperin}}]{Strand:2010gz}%
  \BibitemOpen
  \bibfield  {author} {\bibinfo {author} {\bibfnamefont {J~D}\ \bibnamefont
  {Strand}}, \bibinfo {author} {\bibfnamefont {D~J}\ \bibnamefont {Bahr}},
  \bibinfo {author} {\bibfnamefont {D~J}\ \bibnamefont {Van~Harlingen}},
  \bibinfo {author} {\bibfnamefont {J~P}\ \bibnamefont {Davis}}, \bibinfo
  {author} {\bibfnamefont {W~J}\ \bibnamefont {Gannon}}, \ and\ \bibinfo
  {author} {\bibfnamefont {W~P}\ \bibnamefont {Halperin}},\ }\bibfield  {title}
  {\enquote {\bibinfo {title} {{The Transition Between Real and Complex
  Superconducting Order Parameter Phases in UPt$_3$}},}\ }\href@noop {}
  {\bibfield  {journal} {\bibinfo  {journal} {Science}\ }\textbf {\bibinfo
  {volume} {328}},\ \bibinfo {pages} {1368--1369} (\bibinfo {year}
  {2010})}\BibitemShut {NoStop}%
\bibitem [{\citenamefont {Signore}\ \emph {et~al.}(1995)\citenamefont
  {Signore}, \citenamefont {Andraka}, \citenamefont {Meisel}, \citenamefont
  {Brown}, \citenamefont {Fisk}, \citenamefont {Giorgi}, \citenamefont {Smith},
  \citenamefont {Gross-Alltag}, \citenamefont {Schuberth},\ and\ \citenamefont
  {Menovsky}}]{Signore:1995hu}%
  \BibitemOpen
  \bibfield  {author} {\bibinfo {author} {\bibfnamefont {P~J~C}\ \bibnamefont
  {Signore}}, \bibinfo {author} {\bibfnamefont {B}~\bibnamefont {Andraka}},
  \bibinfo {author} {\bibfnamefont {M~W}\ \bibnamefont {Meisel}}, \bibinfo
  {author} {\bibfnamefont {S~E}\ \bibnamefont {Brown}}, \bibinfo {author}
  {\bibfnamefont {Z}~\bibnamefont {Fisk}}, \bibinfo {author} {\bibfnamefont
  {A~L}\ \bibnamefont {Giorgi}}, \bibinfo {author} {\bibfnamefont {J~L}\
  \bibnamefont {Smith}}, \bibinfo {author} {\bibfnamefont {F}~\bibnamefont
  {Gross-Alltag}}, \bibinfo {author} {\bibfnamefont {E~A}\ \bibnamefont
  {Schuberth}}, \ and\ \bibinfo {author} {\bibfnamefont {A~A}\ \bibnamefont
  {Menovsky}},\ }\bibfield  {title} {\enquote {\bibinfo {title} {{Inductive
  measurements of UPt$_3$ in the superconducting state}},}\ }\href@noop {}
  {\bibfield  {journal} {\bibinfo  {journal} {Phys. Rev. B}\ }\textbf {\bibinfo
  {volume} {52}},\ \bibinfo {pages} {4446--4461} (\bibinfo {year}
  {1995})}\BibitemShut {NoStop}%
\bibitem [{\citenamefont {Sch{\"o}ttl}\ \emph {et~al.}(1999)\citenamefont
  {Sch{\"o}ttl}, \citenamefont {Schuberth}, \citenamefont {Flachbart},
  \citenamefont {Kycia}, \citenamefont {Hong}, \citenamefont {Seidman},
  \citenamefont {Halperin}, \citenamefont {Hufnagl},\ and\ \citenamefont
  {Bucher}}]{Schottl:1999ks}%
  \BibitemOpen
  \bibfield  {author} {\bibinfo {author} {\bibfnamefont {S}~\bibnamefont
  {Sch{\"o}ttl}}, \bibinfo {author} {\bibfnamefont {E~A}\ \bibnamefont
  {Schuberth}}, \bibinfo {author} {\bibfnamefont {K}~\bibnamefont {Flachbart}},
  \bibinfo {author} {\bibfnamefont {J~B}\ \bibnamefont {Kycia}}, \bibinfo
  {author} {\bibfnamefont {J~I}\ \bibnamefont {Hong}}, \bibinfo {author}
  {\bibfnamefont {D~N}\ \bibnamefont {Seidman}}, \bibinfo {author}
  {\bibfnamefont {W~P}\ \bibnamefont {Halperin}}, \bibinfo {author}
  {\bibfnamefont {J}~\bibnamefont {Hufnagl}}, \ and\ \bibinfo {author}
  {\bibfnamefont {E}~\bibnamefont {Bucher}},\ }\bibfield  {title} {\enquote
  {\bibinfo {title} {{Anisotropic dc Magnetization of Superconducting UPt$_3$
  and Antiferromagnetic Ordering Below 20 mK}},}\ }\href@noop {} {\bibfield
  {journal} {\bibinfo  {journal} {Phys. Rev. Lett.}\ }\textbf {\bibinfo
  {volume} {82}},\ \bibinfo {pages} {2378--2381} (\bibinfo {year}
  {1999})}\BibitemShut {NoStop}%
\bibitem [{\citenamefont {Gannon}\ \emph {et~al.}(2015)\citenamefont {Gannon},
  \citenamefont {Halperin}, \citenamefont {Rastovski}, \citenamefont
  {Schlesinger}, \citenamefont {Hlevyack}, \citenamefont {Eskildsen},
  \citenamefont {Vorontsov}, \citenamefont {Gavilano}, \citenamefont {Gasser},\
  and\ \citenamefont {Nagy}}]{Gannon:2015ct}%
  \BibitemOpen
  \bibfield  {author} {\bibinfo {author} {\bibfnamefont {W~J}\ \bibnamefont
  {Gannon}}, \bibinfo {author} {\bibfnamefont {W~P}\ \bibnamefont {Halperin}},
  \bibinfo {author} {\bibfnamefont {C}~\bibnamefont {Rastovski}}, \bibinfo
  {author} {\bibfnamefont {K~J}\ \bibnamefont {Schlesinger}}, \bibinfo {author}
  {\bibfnamefont {J}~\bibnamefont {Hlevyack}}, \bibinfo {author} {\bibfnamefont
  {M~R}\ \bibnamefont {Eskildsen}}, \bibinfo {author} {\bibfnamefont {A~B}\
  \bibnamefont {Vorontsov}}, \bibinfo {author} {\bibfnamefont {J}~\bibnamefont
  {Gavilano}}, \bibinfo {author} {\bibfnamefont {U}~\bibnamefont {Gasser}}, \
  and\ \bibinfo {author} {\bibfnamefont {G}~\bibnamefont {Nagy}},\ }\bibfield
  {title} {\enquote {\bibinfo {title} {{Nodal gap structure and order parameter
  symmetry of the unconventional superconductor UPt$_3$}},}\ }\href@noop {}
  {\bibfield  {journal} {\bibinfo  {journal} {New J. Phys.}\ }\textbf {\bibinfo
  {volume} {17}},\ \bibinfo {pages} {023041} (\bibinfo {year}
  {2015})}\BibitemShut {NoStop}%
\bibitem [{\citenamefont {Kycia}\ \emph {et~al.}(1998)\citenamefont {Kycia},
  \citenamefont {Hong}, \citenamefont {Graf}, \citenamefont {Sauls},
  \citenamefont {Seidman},\ and\ \citenamefont {Halperin}}]{Kycia:1998ui}%
  \BibitemOpen
  \bibfield  {author} {\bibinfo {author} {\bibfnamefont {J~B}\ \bibnamefont
  {Kycia}}, \bibinfo {author} {\bibfnamefont {J~I}\ \bibnamefont {Hong}},
  \bibinfo {author} {\bibfnamefont {M~J}\ \bibnamefont {Graf}}, \bibinfo
  {author} {\bibfnamefont {J~A}\ \bibnamefont {Sauls}}, \bibinfo {author}
  {\bibfnamefont {D~N}\ \bibnamefont {Seidman}}, \ and\ \bibinfo {author}
  {\bibfnamefont {W~P}\ \bibnamefont {Halperin}},\ }\bibfield  {title}
  {\enquote {\bibinfo {title} {{Suppression of superconductivity in UPt$_3$
  single crystals}},}\ }\href@noop {} {\bibfield  {journal} {\bibinfo
  {journal} {Phys. Rev. B}\ }\textbf {\bibinfo {volume} {58}},\ \bibinfo
  {pages} {R603--R606} (\bibinfo {year} {1998})}\BibitemShut {NoStop}%
\bibitem [{\citenamefont {Eskildsen}\ \emph {et~al.}(2016)\citenamefont
  {Eskildsen}, \citenamefont {Avers}, \citenamefont {Dewhurst}, \citenamefont
  {Gannon}, \citenamefont {Halperin},\ and\ \citenamefont {White}}]{5-42-402}%
  \BibitemOpen
  \bibfield  {author} {\bibinfo {author} {\bibfnamefont {M~R}\ \bibnamefont
  {Eskildsen}}, \bibinfo {author} {\bibfnamefont {K}~\bibnamefont {Avers}},
  \bibinfo {author} {\bibfnamefont {C}~\bibnamefont {Dewhurst}}, \bibinfo
  {author} {\bibfnamefont {W}~\bibnamefont {Gannon}}, \bibinfo {author}
  {\bibfnamefont {W~P}\ \bibnamefont {Halperin}}, \ and\ \bibinfo {author}
  {\bibfnamefont {J}~\bibnamefont {White}},\ }\href
  {http://dx.doi.org/10.5291/ILL-DATA.5-42-402} {}\bibinfo {howpublished}
  {{Chiral Effects on the Vortex Lattice in UPt$_3$, Institut Laue-Langevin
  (ILL), Grenoble (France), doi: 10.5291/ILL-DATA.5-42-402}} (\bibinfo {year}
  {2016})\BibitemShut {NoStop}%
\bibitem [{\citenamefont {M\"uhlbauer}\ \emph {et~al.}()\citenamefont
  {M\"uhlbauer}, \citenamefont {Honecker}, \citenamefont {P\'erigo},
  \citenamefont {Bergner}, \citenamefont {Disch}, \citenamefont {Heinemann},
  \citenamefont {Erokhin}, \citenamefont {Berkov}, \citenamefont {Leighton},
  \citenamefont {Eskildsen},\ and\ \citenamefont {Michels}}]{MuhlbauerRMP}%
  \BibitemOpen
  \bibfield  {author} {\bibinfo {author} {\bibfnamefont {S}~\bibnamefont
  {M\"uhlbauer}}, \bibinfo {author} {\bibfnamefont {D}~\bibnamefont
  {Honecker}}, \bibinfo {author} {\bibfnamefont {\'E~A}\ \bibnamefont
  {P\'erigo}}, \bibinfo {author} {\bibfnamefont {F}~\bibnamefont {Bergner}},
  \bibinfo {author} {\bibfnamefont {S}~\bibnamefont {Disch}}, \bibinfo {author}
  {\bibfnamefont {A}~\bibnamefont {Heinemann}}, \bibinfo {author}
  {\bibfnamefont {S}~\bibnamefont {Erokhin}}, \bibinfo {author} {\bibfnamefont
  {D}~\bibnamefont {Berkov}}, \bibinfo {author} {\bibfnamefont {C}~\bibnamefont
  {Leighton}}, \bibinfo {author} {\bibfnamefont {M~R}\ \bibnamefont
  {Eskildsen}}, \ and\ \bibinfo {author} {\bibfnamefont {A}~\bibnamefont
  {Michels}},\ }\bibfield  {title} {\enquote {\bibinfo {title} {{Magnetic
  small-angle neutron scattering}},}\ }\href@noop {} {\bibinfo  {journal} {to
  appear in Rev. Mod. Phys}\ }\BibitemShut {NoStop}%
\bibitem [{SM()}]{SM}%
  \BibitemOpen
\bibfield  {journal} {  }\href@noop {} {}\bibinfo {howpublished} {Supplemental
  Material}\BibitemShut {NoStop}%
\bibitem [{\citenamefont {Huxley}\ \emph {et~al.}(2000)\citenamefont {Huxley},
  \citenamefont {Rodi{\`e}re}, \citenamefont {Paul}, \citenamefont {van Dijk},
  \citenamefont {Cubitt},\ and\ \citenamefont {Flouquet}}]{Huxley:2000aa}%
  \BibitemOpen
  \bibfield  {author} {\bibinfo {author} {\bibfnamefont {A}~\bibnamefont
  {Huxley}}, \bibinfo {author} {\bibfnamefont {P}~\bibnamefont {Rodi{\`e}re}},
  \bibinfo {author} {\bibfnamefont {D~McK}\ \bibnamefont {Paul}}, \bibinfo
  {author} {\bibfnamefont {N}~\bibnamefont {van Dijk}}, \bibinfo {author}
  {\bibfnamefont {R}~\bibnamefont {Cubitt}}, \ and\ \bibinfo {author}
  {\bibfnamefont {J}~\bibnamefont {Flouquet}},\ }\bibfield  {title} {\enquote
  {\bibinfo {title} {{Realignment of the flux-line lattice by a change in the
  symmetry of superconductivity in UPt$3$}},}\ }\href@noop {} {\bibfield
  {journal} {\bibinfo  {journal} {Nature}\ }\textbf {\bibinfo {volume} {406}},\
  \bibinfo {pages} {160--164} (\bibinfo {year} {2000})}\BibitemShut {NoStop}%
\bibitem [{\citenamefont {Yaron}\ \emph {et~al.}(1997)\citenamefont {Yaron},
  \citenamefont {Gammel}, \citenamefont {Boebinger}, \citenamefont {Aeppli},
  \citenamefont {Schiffer}, \citenamefont {Bucher}, \citenamefont {Bishop},
  \citenamefont {Broholm},\ and\ \citenamefont {Mortensen}}]{Yaron:1997wx}%
  \BibitemOpen
  \bibfield  {author} {\bibinfo {author} {\bibfnamefont {U}~\bibnamefont
  {Yaron}}, \bibinfo {author} {\bibfnamefont {P~L}\ \bibnamefont {Gammel}},
  \bibinfo {author} {\bibfnamefont {G~S}\ \bibnamefont {Boebinger}}, \bibinfo
  {author} {\bibfnamefont {G}~\bibnamefont {Aeppli}}, \bibinfo {author}
  {\bibfnamefont {P}~\bibnamefont {Schiffer}}, \bibinfo {author} {\bibfnamefont
  {E}~\bibnamefont {Bucher}}, \bibinfo {author} {\bibfnamefont {D~J}\
  \bibnamefont {Bishop}}, \bibinfo {author} {\bibfnamefont {C}~\bibnamefont
  {Broholm}}, \ and\ \bibinfo {author} {\bibfnamefont {K}~\bibnamefont
  {Mortensen}},\ }\bibfield  {title} {\enquote {\bibinfo {title} {{Small Angle
  Neutron Scattering Studies of the Vortex Lattice in the UPt$_3$ Mixed State:
  Direct Structural Evidence for the $B$ to $C$ Transition}},}\ }\href@noop {}
  {\bibfield  {journal} {\bibinfo  {journal} {Phys. Rev. Lett.}\ }\textbf
  {\bibinfo {volume} {78}},\ \bibinfo {pages} {3185--3188} (\bibinfo {year}
  {1997})}\BibitemShut {NoStop}%
\bibitem [{\citenamefont {Eskildsen}\ \emph {et~al.}(1997)\citenamefont
  {Eskildsen}, \citenamefont {Gammel}, \citenamefont {Barber}, \citenamefont
  {Ramirez}, \citenamefont {Bishop}, \citenamefont {Andersen}, \citenamefont
  {Mortensen}, \citenamefont {Bolle}, \citenamefont {Lieber},\ and\
  \citenamefont {Canfield}}]{Eskildsen:1997ab}%
  \BibitemOpen
  \bibfield  {author} {\bibinfo {author} {\bibfnamefont {M~R}\ \bibnamefont
  {Eskildsen}}, \bibinfo {author} {\bibfnamefont {P~L}\ \bibnamefont {Gammel}},
  \bibinfo {author} {\bibfnamefont {B~P}\ \bibnamefont {Barber}}, \bibinfo
  {author} {\bibfnamefont {A~P}\ \bibnamefont {Ramirez}}, \bibinfo {author}
  {\bibfnamefont {D~J}\ \bibnamefont {Bishop}}, \bibinfo {author}
  {\bibfnamefont {N~H}\ \bibnamefont {Andersen}}, \bibinfo {author}
  {\bibfnamefont {K}~\bibnamefont {Mortensen}}, \bibinfo {author}
  {\bibfnamefont {C~A}\ \bibnamefont {Bolle}}, \bibinfo {author} {\bibfnamefont
  {C~M}\ \bibnamefont {Lieber}}, \ and\ \bibinfo {author} {\bibfnamefont {P~C}\
  \bibnamefont {Canfield}},\ }\bibfield  {title} {\enquote {\bibinfo {title}
  {{Structural stability of the square flux line lattice in YNi$_2$B$_2$C and
  LuNi$_2$B$_2$C studied with small angle neutron scattering}},}\ }\href@noop
  {} {\bibfield  {journal} {\bibinfo  {journal} {Phys. Rev. Lett.}\ }\textbf
  {\bibinfo {volume} {79}},\ \bibinfo {pages} {487--490} (\bibinfo {year}
  {1997})}\BibitemShut {NoStop}%
\bibitem [{\citenamefont {Tokuyasu}\ and\ \citenamefont
  {Sauls}(1990)}]{Tokuyasu:1990tk}%
  \BibitemOpen
  \bibfield  {author} {\bibinfo {author} {\bibfnamefont {T~A}\ \bibnamefont
  {Tokuyasu}}\ and\ \bibinfo {author} {\bibfnamefont {J~A}\ \bibnamefont
  {Sauls}},\ }\bibfield  {title} {\enquote {\bibinfo {title} {{Stability of
  doubly quantized vortices in unconventional superconductors}},}\ }\href@noop
  {} {\bibfield  {journal} {\bibinfo  {journal} {Physica B}\ }\textbf {\bibinfo
  {volume} {165--166}},\ \bibinfo {pages} {347--348} (\bibinfo {year}
  {1990})}\BibitemShut {NoStop}%
\bibitem [{\citenamefont {Sauls}\ and\ \citenamefont
  {Eschrig}(2009)}]{Sauls:2009dw}%
  \BibitemOpen
  \bibfield  {author} {\bibinfo {author} {\bibfnamefont {J~A}\ \bibnamefont
  {Sauls}}\ and\ \bibinfo {author} {\bibfnamefont {M}~\bibnamefont {Eschrig}},\
  }\bibfield  {title} {\enquote {\bibinfo {title} {{Vortices in chiral,
  spin-triplet superconductors and superfluids}},}\ }\href@noop {} {\bibfield
  {journal} {\bibinfo  {journal} {New J. Phys.}\ }\textbf {\bibinfo {volume}
  {11}},\ \bibinfo {pages} {075008} (\bibinfo {year} {2009})}\BibitemShut
  {NoStop}%
\bibitem [{\citenamefont {Ichioka}\ \emph {et~al.}(2012)\citenamefont
  {Ichioka}, \citenamefont {Machida},\ and\ \citenamefont
  {Sauls}}]{Ichioka:2012wy}%
  \BibitemOpen
  \bibfield  {author} {\bibinfo {author} {\bibfnamefont {M}~\bibnamefont
  {Ichioka}}, \bibinfo {author} {\bibfnamefont {K}~\bibnamefont {Machida}}, \
  and\ \bibinfo {author} {\bibfnamefont {J~A}\ \bibnamefont {Sauls}},\
  }\bibfield  {title} {\enquote {\bibinfo {title} {{Vortex States of Chiral
  $p$-wave Superconductors}},}\ }\href@noop {} {\bibfield  {journal} {\bibinfo
  {journal} {J. Phys.: Conf. Ser.}\ }\textbf {\bibinfo {volume} {400}},\
  \bibinfo {pages} {022031} (\bibinfo {year} {2012})}\BibitemShut {NoStop}%
\bibitem [{\citenamefont {Volovik}(1988)}]{Volovik:1988ul}%
  \BibitemOpen
  \bibfield  {author} {\bibinfo {author} {\bibfnamefont {G~E}\ \bibnamefont
  {Volovik}},\ }\bibfield  {title} {\enquote {\bibinfo {title} {{On the vortex
  lattice transition in heavy-fermionic UPt$_3$}},}\ }\href@noop {} {\bibfield
  {journal} {\bibinfo  {journal} {J. Phys. C: Solid St. Phys.}\ }\textbf
  {\bibinfo {volume} {21}},\ \bibinfo {pages} {L221--L224} (\bibinfo {year}
  {1988})}\BibitemShut {NoStop}%
\bibitem [{\citenamefont {Tsutsumi}\ \emph {et~al.}(2012)\citenamefont
  {Tsutsumi}, \citenamefont {Machida}, \citenamefont {Ohmi},\ and\
  \citenamefont {Ozaki}}]{Tsutsumi:2012hj}%
  \BibitemOpen
  \bibfield  {author} {\bibinfo {author} {\bibfnamefont {Y}~\bibnamefont
  {Tsutsumi}}, \bibinfo {author} {\bibfnamefont {K}~\bibnamefont {Machida}},
  \bibinfo {author} {\bibfnamefont {T}~\bibnamefont {Ohmi}}, \ and\ \bibinfo
  {author} {\bibfnamefont {M-a}\ \bibnamefont {Ozaki}},\ }\bibfield  {title}
  {\enquote {\bibinfo {title} {{A Spin Triplet Superconductor UPt$_3$}},}\
  }\href@noop {} {\bibfield  {journal} {\bibinfo  {journal} {J. Phys. Soc.
  Jpn.}\ }\textbf {\bibinfo {volume} {81}},\ \bibinfo {pages} {074717}
  (\bibinfo {year} {2012})}\BibitemShut {NoStop}%
\bibitem [{\citenamefont {Takamatsu}\ and\ \citenamefont
  {Yanase}(2015)}]{Takamatsu:2015cv}%
  \BibitemOpen
  \bibfield  {author} {\bibinfo {author} {\bibfnamefont {S}~\bibnamefont
  {Takamatsu}}\ and\ \bibinfo {author} {\bibfnamefont {Y}~\bibnamefont
  {Yanase}},\ }\bibfield  {title} {\enquote {\bibinfo {title} {{Chiral
  superconductivity in nematic states}},}\ }\href@noop {} {\bibfield  {journal}
  {\bibinfo  {journal} {Phys. Rev. B}\ }\textbf {\bibinfo {volume} {91}},\
  \bibinfo {pages} {054504} (\bibinfo {year} {2015})}\BibitemShut {NoStop}%
\bibitem [{\citenamefont {Kogan}\ \emph {et~al.}(1997)\citenamefont {Kogan},
  \citenamefont {Miranovic}, \citenamefont {Dobrosavlevic-Grujic},
  \citenamefont {Pickett},\ and\ \citenamefont {Christen}}]{Kogan:1997wy}%
  \BibitemOpen
  \bibfield  {author} {\bibinfo {author} {\bibfnamefont {V~G}\ \bibnamefont
  {Kogan}}, \bibinfo {author} {\bibfnamefont {P}~\bibnamefont {Miranovic}},
  \bibinfo {author} {\bibfnamefont {Lj}~\bibnamefont {Dobrosavlevic-Grujic}},
  \bibinfo {author} {\bibfnamefont {W~E}\ \bibnamefont {Pickett}}, \ and\
  \bibinfo {author} {\bibfnamefont {D~K}\ \bibnamefont {Christen}},\ }\bibfield
   {title} {\enquote {\bibinfo {title} {{Vortex Lattices in Cubic
  Superconductors}},}\ }\href@noop {} {\bibfield  {journal} {\bibinfo
  {journal} {Phys. Rev. Lett.}\ }\textbf {\bibinfo {volume} {79}},\ \bibinfo
  {pages} {741--744} (\bibinfo {year} {1997})}\BibitemShut {NoStop}%
\bibitem [{\citenamefont {Champel}\ and\ \citenamefont
  {Mineev}(2001)}]{Champel:2001kf}%
  \BibitemOpen
  \bibfield  {author} {\bibinfo {author} {\bibfnamefont {T}~\bibnamefont
  {Champel}}\ and\ \bibinfo {author} {\bibfnamefont {V~P}\ \bibnamefont
  {Mineev}},\ }\bibfield  {title} {\enquote {\bibinfo {title} {{Theory of
  Equilibrium Flux Lattice in UPt$_3$ under Magnetic Field Parallel to
  Hexagonal Crystal Axis}},}\ }\href@noop {} {\bibfield  {journal} {\bibinfo
  {journal} {Phys. Rev. Lett.}\ }\textbf {\bibinfo {volume} {86}},\ \bibinfo
  {pages} {4903--4906} (\bibinfo {year} {2001})}\BibitemShut {NoStop}%
\bibitem [{\citenamefont {Affleck}\ \emph {et~al.}(1997)\citenamefont
  {Affleck}, \citenamefont {Franz},\ and\ \citenamefont
  {Sharifzadeh~A}}]{Affleck:1997bg}%
  \BibitemOpen
  \bibfield  {author} {\bibinfo {author} {\bibfnamefont {I}~\bibnamefont
  {Affleck}}, \bibinfo {author} {\bibfnamefont {M}~\bibnamefont {Franz}}, \
  and\ \bibinfo {author} {\bibfnamefont {M~H}\ \bibnamefont {Sharifzadeh~A}},\
  }\bibfield  {title} {\enquote {\bibinfo {title} {{Generalized London free
  energy for high-$T_c$ vortex lattices}},}\ }\href@noop {} {\bibfield
  {journal} {\bibinfo  {journal} {Phys. Rev. B}\ }\textbf {\bibinfo {volume}
  {55}},\ \bibinfo {pages} {R704--R707} (\bibinfo {year} {1997})}\BibitemShut
  {NoStop}%
\bibitem [{\citenamefont {Tokuyasu}\ \emph {et~al.}(1990)\citenamefont
  {Tokuyasu}, \citenamefont {Hess},\ and\ \citenamefont
  {Sauls}}]{Tokuyasu:1990ui}%
  \BibitemOpen
  \bibfield  {author} {\bibinfo {author} {\bibfnamefont {T~A}\ \bibnamefont
  {Tokuyasu}}, \bibinfo {author} {\bibfnamefont {D~W}\ \bibnamefont {Hess}}, \
  and\ \bibinfo {author} {\bibfnamefont {J~A}\ \bibnamefont {Sauls}},\
  }\bibfield  {title} {\enquote {\bibinfo {title} {{Vortex states in an
  unconventional superconductor and the mixed phases of UPt$_3$}},}\
  }\href@noop {} {\bibfield  {journal} {\bibinfo  {journal} {Phys. Rev. B}\
  }\textbf {\bibinfo {volume} {41}},\ \bibinfo {pages} {8891--8903} (\bibinfo
  {year} {1990})}\BibitemShut {NoStop}%
\bibitem [{\citenamefont {Ichioka}\ and\ \citenamefont
  {Machida}(2002)}]{Ichioka:2002go}%
  \BibitemOpen
  \bibfield  {author} {\bibinfo {author} {\bibfnamefont {M}~\bibnamefont
  {Ichioka}}\ and\ \bibinfo {author} {\bibfnamefont {K}~\bibnamefont
  {Machida}},\ }\bibfield  {title} {\enquote {\bibinfo {title} {{Field
  dependence of the vortex structure in chiral $p$-wave superconductors}},}\
  }\href@noop {} {\bibfield  {journal} {\bibinfo  {journal} {Phys. Rev. B}\
  }\textbf {\bibinfo {volume} {65}},\ \bibinfo {pages} {224517} (\bibinfo
  {year} {2002})}\BibitemShut {NoStop}%
\bibitem [{\citenamefont {Nomoto}\ and\ \citenamefont
  {Ikeda}(2016)}]{Nomoto:2016ko}%
  \BibitemOpen
  \bibfield  {author} {\bibinfo {author} {\bibfnamefont {T}~\bibnamefont
  {Nomoto}}\ and\ \bibinfo {author} {\bibfnamefont {H}~\bibnamefont {Ikeda}},\
  }\bibfield  {title} {\enquote {\bibinfo {title} {{Exotic Multigap Structure
  in UPt$_3$ Unveiled by a First-Principles Analysis}},}\ }\href@noop {}
  {\bibfield  {journal} {\bibinfo  {journal} {Phys. Rev. Lett.}\ }\textbf
  {\bibinfo {volume} {117}},\ \bibinfo {pages} {217002} (\bibinfo {year}
  {2016})}\BibitemShut {NoStop}%
\bibitem [{\citenamefont {Thuneberg}(1986)}]{Thuneberg:1986ix}%
  \BibitemOpen
  \bibfield  {author} {\bibinfo {author} {\bibfnamefont {E~V}\ \bibnamefont
  {Thuneberg}},\ }\bibfield  {title} {\enquote {\bibinfo {title}
  {{Identification of vortices in superfluid $^{3}$He-\emph{B}}},}\ }\href@noop
  {} {\bibfield  {journal} {\bibinfo  {journal} {Phys. Rev. Lett.}\ }\textbf
  {\bibinfo {volume} {56}},\ \bibinfo {pages} {359--362} (\bibinfo {year}
  {1986})}\BibitemShut {NoStop}%
\end{thebibliography}
\end{document}


\title{Supplemental Material: Vortex lattices and broken time reversal symmetry in the topological superconductor {\UPt}}

\author{K.~E.~Avers}
\affiliation{Department of Physics and Astronomy, Northwestern University, Evanston, IL 60208, USA}

\author{W.~J.~Gannon}
\affiliation{Department of Physics and Astronomy, Northwestern University, Evanston, IL 60208, USA}

\author{S.~J.~Kuhn}
\affiliation{Department of Physics, University of Notre Dame, Notre Dame, IN 46556, USA}

\author{W.~P.~Halperin}
\affiliation{Department of Physics and Astronomy, Northwestern University, Evanston, IL 60208, USA}

\author{J.~A.~Sauls}
\affiliation{Center for Applied Physics \& Superconducting Technologies, Northwestern University, Evanston, IL 60208, USA}

\author{L.~DeBeer-Schmitt}
\affiliation{Oak Ridge National Laboratory, Oak Ridge, TN 37831, USA}

\author{C.~D.~Dewhurst}
\affiliation{Institut Laue-Langevin, 71 avenue des Martyrs, CS 20156, F-38042 Grenoble cedex 9, France}

\author{J.~Gavilano}
\affiliation{Laboratory for Neutron Scattering, Paul Scherrer Institute, CH-5232 Villigen, Switzerland}

\author{G.~Nagy}
\affiliation{Laboratory for Neutron Scattering, Paul Scherrer Institute, CH-5232 Villigen, Switzerland}

\author{U.~Gasser}
\affiliation{Laboratory for Neutron Scattering, Paul Scherrer Institute, CH-5232 Villigen, Switzerland}

\author{M.~R.~Eskildsen}
\affiliation{Department of Physics, University of Notre Dame, Notre Dame, IN 46556, USA}

\maketitle

\section*{Field dependence of VL Bragg peak splitting}
The field dependence of the VL Bragg peak splitting in the B phase is well described by the empirical function
\begin{equation}
\omega(H)=A \left(\frac{H}{H_0}\right)^{\alpha}\;\left(1-\frac{H}{H_0} \right)^{\beta}
\,.
\end{equation}
Fitting the field reduced and field reversed data separately, using a constant $H_0 = 1.7$~T, yield exponents $\alpha_{\text{Red.}} = 0.65 \pm 0.06$, $\beta_{\text{Red.}} = 0.89 \pm 0.07$, $\alpha_{\text{Rev.}} = 0.62 \pm 0.06$ and $\beta_{\text{Rev.}} = 0.70 \pm 0.06$. The results of the fits are shown in Fig.~3(a) of the main text. For both field histories one finds $\alpha \neq \beta$, as expected if the non-monotonic field dependence of $\omega$ is due to a competition between two different mechanisms. The value of $H_0$ was chosen to obtain a good fit to the data. However, this is not the B-C phase transition field since $\omega$ does not go to zero in the C phase. The difference between the fits to the field reduced and field reversed data is shown in Fig.~3(b) of the main text.

\section*{Field dependence of scattered intensity}
The scattered intensity is related to the form factor $F(Q)$, given by the Fourier transform of the magnetic field modulation due to the VL. The simplest expression for $F(Q)$ is given by the London model multiplied by a Gaussian cut-off term that accounts for the finite size of vortex cores:
\begin{equation}
  F(Q) = \frac{B}{1 + \lambda_{ab}^2 Q^2} \, e^{-c \xi_{ab}^2 Q^2}.
  \label{FF1}
\end{equation}
Here $\lambda_{ab}$ and $\xi_{ab}$ are respectively penetration depth and coherence length in the screening current plane ($ab$-plane), and $c$ is a constant of order unity (typically between $\tfrac{1}{4}$ and 2))~\cite{MuhlbauerRMP}. Previous SANS measurements yielded a penetration depth $\lambda_{ab} = 680$~nm ~\cite{Gannon:2015ct}, and consequently $(\lambda_{ab} Q)^2 \gg 1$ for the entire range of applied magnetic fields in this work. As a result the expression for the form factor simplifies to
\begin{equation}
  F(Q) = \frac{\sqrt{3} \Phi_0}{8 \pi^2 \lambda_{ab}^2} \, e^{-c \xi_{ab}^2 Q^2},
  \label{FF2}
\end{equation}
using the relation between $B$ and $Q$ from Eq.~(\ref{BQ}). The only field dependence is thus through the core cut-off term, and the form factor is thus expected to decrease exponentially with increasing field as $B \propto Q^2$.

The total (integrated) scattered intensity $\Iint$ of a VL Bragg peak is related to the form factor by~\cite{MuhlbauerRMP}
\begin{equation}
\Iint \propto \frac{| F(Q)|^2}{Q},
\end{equation}
and $\Iint \times Q$ should decrease exponentially with increasing field. This is indeed observed in Fig.~4(b) of the main text for fields between $0.8$~T and $1.5$~T. A fit to data in this field range yields $\xi_{ab} = 9.6 \pm 0.7$~nm,  with a commonly used $c = \tfrac{1}{2}$~\cite{MuhlbauerRMP}. A proper determination of $\Iint$ requires a measurement of the scattered intensity as the VL Bragg peak is rotated through the Ewald sphere in a so-called rocking curve. In the present case only the peak intensity ($I_0$) was measured. As long as the width of the Bragg peak perpendicular to the detector plane is constant $\Iint \propto I_0$. At both low and high fields $I_0 \times Q$ falls below the extrapolation of the intermediate field fit, indicating a broadening of the rocking curves corresponding to a disordering of the VL. An alternative fitting of the low field behavior by Eq.~(\ref{FF1}) using the penetration depth as a free parameter is unacceptable, as it yields a very small value for  $\lambda_{ab} = 6.1$~nm. This is inconsistent with the result obtained from full rocking curves which allow a determination of the VL form factor on an absolute scale, $\lambda_{ab} = 680$~nm~\cite{Gannon:2015ct}.

\section*{Magnetic induction}
\begin{figure}[b]
\includegraphics{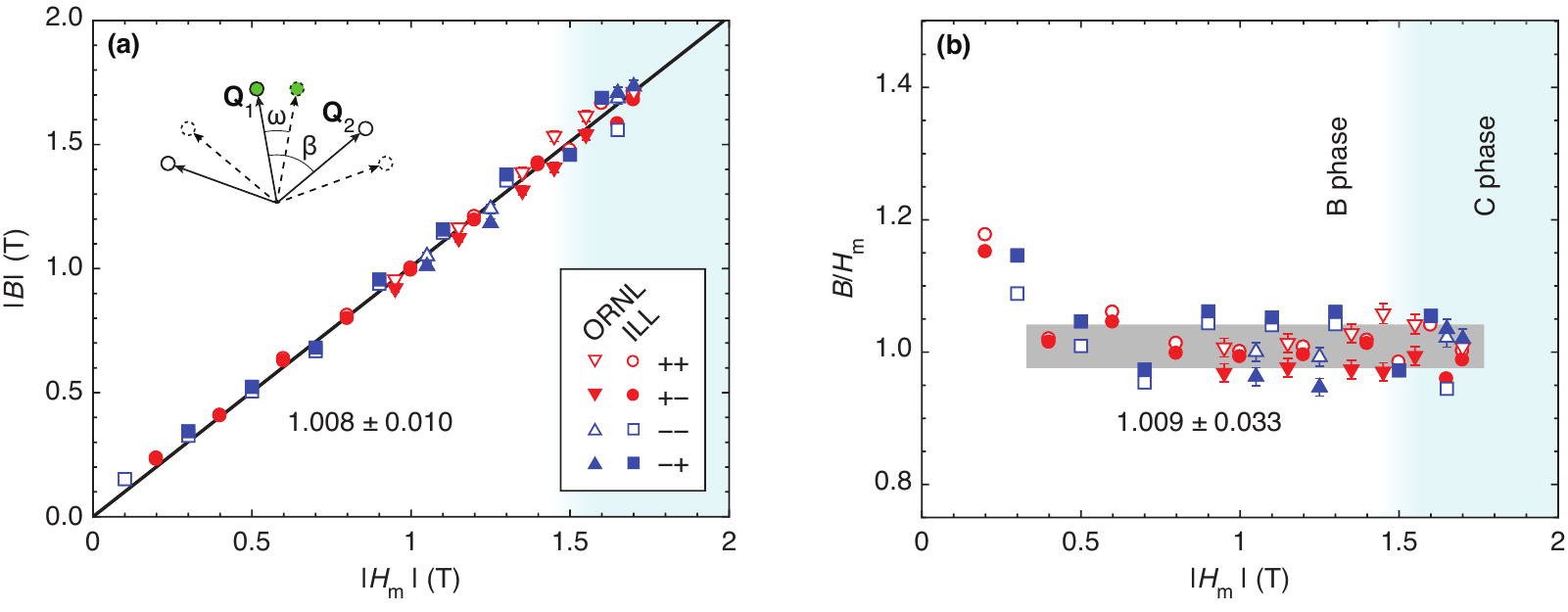}
\caption{\label{BvsH}
  (Color)
  (a) Magnetic induction (vortex density), determined from the magnitude of the VL scattering vectors, versus applied magnetic field.
  The inset shows VL Bragg peaks, their corresponding scattering vectors (${\bf Q}_{1/2}$), and the opening angle ($\beta$).
  Peaks associated with the two different VL domain orientations are indicated by full and dashed lines respectively.
  Only Bragg peaks indicated by solid symbols were imaged by SANS [Fig.~2 of main text].
  The line is a a linear fit to the data. (b) Ratio of magnetic induction to applied field ($B/H$).
  The center and width of the gray bar indicates the average and standard deviation of the values for fields $\geq 0.4$~T.}
\end{figure}

The magnetic induction, $B$, is directly related to the vortex density as each vortex carries one quantum of magnetic flux, $\Phi_0=h/2e=2069$~T~nm$^2$. Since the VL is rotated away from the high-symmetry crystalline directions ($\bf{a}, \bf{a}^*$), two domains were observed at all measured fields. Using two scattering vectors belonging to the same VL domain orientation, illustrated in the inset to Fig.~\ref{BvsH}(a), the magnetic induction is given by
\begin{equation}
B = \frac{\fq}{4\pi^2} | {\bf Q}_1 \times {\bf Q}_2 |.
\end{equation}
For a triangular VL (opening angle $\beta = 60^{\circ}$) the expression for the magnetic induction simplifies to
\begin{equation}
B = \frac{\sqrt{3}}{2} \left(\frac{Q}{2\pi}\right)^2 \fq
\,.
\label{BQ}
\end{equation}
Using Eq.~(\ref{BQ}) and the magnitude of the VL scattering vectors $Q$, determined from two-Gaussian fits to the diffraction patterns, we obtain the magnetic induction as a function of applied field shown in Fig.~\ref{BvsH}(a). A linear fit to the data yields a slope of $1.008 \pm 0.010$, confirming the triangular VL symmetry. Fitting the field reduced ($++$,$--$) and field reversed ($+-$,$-+$) results separately yields slopes of $1.014 \pm 0.012$ and $1.001 \pm 0.015$ respectively. Within the precision of the measurements these two results agree, and there is thus no observable difference in the vortex density for the two field histories.
Figure~\ref{BvsH}(b) shows the ratio of magnetic induction to applied field. At low fields there is a slight deviation from unity, but above 0.4~T we find that $B/H = 1.009 \pm 0.033$. There is no measurable deviation from this value upon entering the C phase. The agreement between $H$ and $B$ also excludes the presence of multiply quantized vortices, which would require the substitution $\Phi_0 \rightarrow n \Phi_0$ in Eq.~(\ref{BQ}).

\section*{Order parameter}

\begin{figure}[h]
  \includegraphics{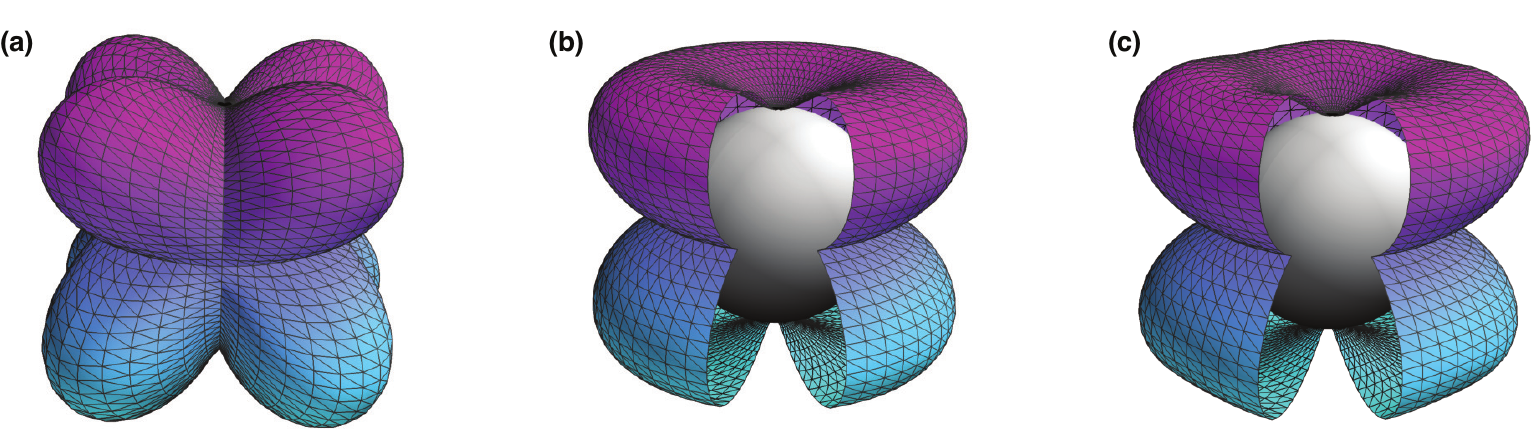}
  \caption{\label{OP}
  (Color)
  Gap amplitudes for the superconducting phases within the $E_{2u}$ model for (a) the C phase, (b) the B phase without the effect of the symmetry breaking field, and (c) the B phase with the anisotropy induced by the symmetry breaking field.}
\end{figure}

For pairing within the $E_{2u}$ representation the structure of the gap amplitudes in the B and C phases are of the form
\begin{equation}
|\Delta(\mathbf p)| \propto \left( \hat{p}_z^2 \left[ ( \hat{p}_x^2 - \hat{p}_y^2)^2 + ((1 - \epsilon) 2 \hat{p}_x \hat{p}_y)^2 \right] \right)^{1/2}.
\label{OPBC}
\end{equation}
Figure~\ref{OP} shows the order parameter for the C phase ($\epsilon=1$), as well as for the B phase, both without ($\epsilon = 0$) and with ($\epsilon = 0.1$) anisotropy induced by the symmetry breaking field, $\epsilon\propto\Delta T_c/T_c\approx 10\%$~\cite{Sauls:1994wd}.

%